# Gulf Stream contribution to recent North Pacific warming

**Yoko Yamagami[1], Hiroaki Tatebe[1], Tsubasa Kohyama[2], Shoichiro Kido[1], Satoru Okajima[3], Yoshiki Komuro[1]**

[1]Japan Agency for Marine-Earth Science and Technology, Yokohama, Japan

[2]Department of Information Sciences, Ochanomizu University, Tokyo, Japan

[3]Institute of Life and Environmental Sciences, University of Tsukuba, Tsukuba, Japan

Corresponding author: Yoko Yamagami (y.yamagami@jamstec.go.jp)

†Research Center for Environmental Modeling and Application, Japan Agency for Marine-Earth Science and Technology, 3173-25 Showamachi, Kanazawaku, Yokohama, Kanagawa 236-0001, Japan

## Abstract

Recent unprecedented ocean warming has produced coherent sea surface temperature (SST) anomalies across the Northern Hemisphere extratropics. While the tropical Pacific is a natural source of North Pacific variability, the influence of the midlatitude North Atlantic has remained poorly understood. Here we show that Gulf Stream SST variability remotely modulates Kuroshio variability, explaining 13% of internal Kuroshio SST variance in climate model simulations, with no significant reverse influence. Positive Gulf Stream SST anomalies excite a Northern Annular Mode (NAM)-like atmospheric circulation response, weakening the Aleutian Low over the North Pacific. The resulting northward shift of the Kuroshio Extension enhances northward warm-water transport and favors positive SST anomalies in the western midlatitude North Pacific. We also show that a high-resolution ocean model is required to properly capture this SST-NAM coupling, suggesting the importance of representations of western boundary currents and associated SST fronts for this interbasin pathway. These findings reveal the Gulf Stream–Kuroshio linkage through which North Atlantic variability has measurably contributed to the recent exceptional North Pacific warming, implying a previously underrecognized source of decadal climate predictability.

## Introduction

The global mean temperature reached record-breaking levels in 2023 and 2024[1–5], intensifying concerns about accelerating impacts of climate change. Global warming has manifested as severe marine heatwaves across global oceans[6] (Fig. 1a,b), with abnormal sea surface temperature (SST) anomalies contributing to extreme summer heatwaves over the mid-latitude regions[7–9]. In addition to anthropogenic global warming[10], the mid-latitude climate system varies internally[11–14], necessitating improved understanding of the physical mechanisms that drive this variability. While internal climate variability in the tropical Pacific has been recognized as a global climate pacemaker[15,16], increasing evidence also highlights the influence of mid-latitude ocean processes on large-scale climate variability[17,18]. However, climate models from the Coupled Model Intercomparison Project Phase 6 (CMIP6) tend to underestimate the potential predictability of the mid-latitude climate[19,20], complicating both the prediction of climate anomalies and the understanding of the mechanisms underlying their variability[21].

The Kuroshio and Gulf Stream, two major western boundary currents, exert a fundamental influence on Northern Hemisphere climate through their sharp SST fronts and active mesoscale eddy fields. SST anomalies and their gradients in these regions modulate not only the marine boundary layer[22] but also deep convection[23], storm track activity[24–26], the North Atlantic Oscillation (NAO)[27,28], and the Northern Hemisphere annular mode (NAM)[29,30]. However, mid-latitude SST anomalies were historically regarded as passive responses to stochastic atmospheric forcing[31], and as a result, inter-basin interactions between the Kuroshio and Gulf Stream have remained relatively unexplored. Notable exceptions include a study that proposed the possibility of dynamical connections between the two western boundary currents based on a simple linear air-sea interaction model[32] and observational data analysis[33]. A climate model simulation further suggested that Kuroshio Extension variability may exhibit a five-year delayed response to Atlantic Multidecadal Variability (AMV)[34]. More recently, high-resolution climate models, combined with observations and a simple theoretical framework, have revealed a boundary current synchronization between the Kuroshio and Gulf Stream[35].

Western boundary current variability is regulated by oceanic Rossby waves and exhibits multiyear thermal memory[14,36], potentially serving as a driver of internal variability in the mid-latitude climate system. Consequently, interactions between the Kuroshio and the Gulf Stream[32–35] could provide a predominant source of seasonal-to-decadal predictability for the Northern Hemisphere climate. However, the causal relationship and dynamical mechanisms linking these current systems have remained poorly understood. To address this knowledge gap, here we conduct a suite of climate simulations in which SST anomalies in the Kuroshio and the Gulf Stream regions are constrained using various climate model configurations. These experiments allow us to clarify the mechanisms of interactions between the Gulf Stream and the Kuroshio, as well as the relevance of the inter-basin interactions to recent North Pacific mid-latitude SST warming. The results reveal a newly identified modulatory role of Gulf Stream SST in decadal Kuroshio variability through NAM-like atmospheric circulation anomalies, and suggest that this pathway has contributed to the recent exceptional ocean warming observed in the North Pacific.

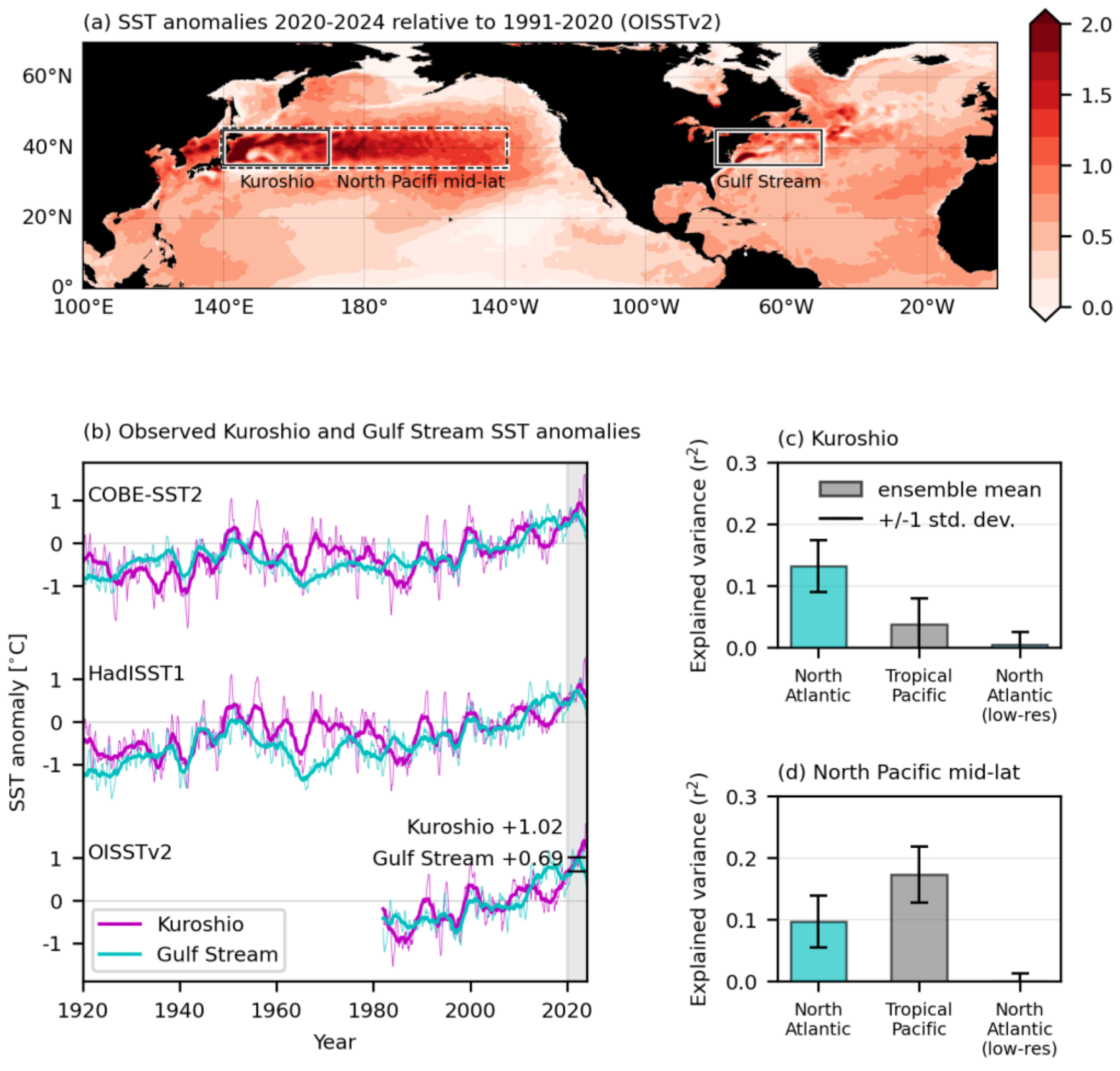

**Figure 1. Abnormally warm sea surface temperature (SST) over the Northern Hemisphere in recent years and simulated contribution of remote forcings to North Pacific.** (a) Observed SST anomalies during 2020-2024, relative to the 1991-2020 annual mean climatology. (b) Observed monthly mean SST anomalies relative to 1991-2020 climatology, area-averaged over the Kuroshio region (140°E–170°E, 35°N–45°N; magenta) and Gulf Stream (50°–80°W, 35°N–45°N; cyan), as indicated by black boxes in (a), based on COBE-SST2, HadISST, and OISSTv2. Thin lines indicate 7-month running mean and thick lines are 3-year running mean. Grey shading indicates the 2020-2024 period. The mean anomaly for OISSTv2 over this period is shown in the bottom. (c) Simulated contributions of North Atlantic mid-

latitude and tropical Pacific variability to SST anomalies area-averaged over the Kuroshio region. North Atlantic mid-latitude contribution simulated by a non-eddying configuration is also shown. Bars indicate the explained variances simulated as the ensemble means, while error bars were estimated as ±1 standard deviations for ensemble members. (d) As in (c), but for entire North Pacific mid-latitude SST (140E-140W, 35N-45N; black dashed box in (a)).

**Modulatory contribution of the Gulf Stream to Kuroshio variability**

To investigate the mutual influence between SST anomalies in the Kuroshio and Gulf Stream regions, we conducted a series of ensemble experiments using the Model for Interdisciplinary Research on Climate version 6 (MIROC6[37]), a participant in CMIP6[38]. The reference simulation (CTL) is a 100-year pre-industrial control simulation with a globally eddy-permitting ocean resolution (Methods). Two ensemble experiments, configured similarly to CTL, constrain SST anomalies in either the North Pacific or North Atlantic using SST anomalies derived from CTL. These experiments, each consisting of ten ensemble members, are referred to as North Pacific-Global Atmosphere (NPGA) and North Atlantic-Global Atmosphere (NAGA), respectively (Methods). In addition, to separate the possible contribution of tropical Pacific variability, we also conducted a Tropical Pacific-Global Atmosphere (POGA) experiment, in which tropical Pacific SST anomalies were constrained in the same manner as NPGA and NAGA (Methods).

Comparative analysis reveals an asymmetric relationship in which low-frequency SST variability in the Kuroshio region (140°E-170°E, 35°N-45°N; left black box in Fig. 1a) is significantly modulated by Gulf Stream SSTs (50°-80°W, 35°N-45°N; right black box in Fig. 1a) (Extended Data 1). In the NAGA ensemble mean, Kuroshio SSTs exhibit a significant correlation with CTL ($r = 0.36$), whereas Gulf Stream SSTs in NPGA show a negligible correlation with CTL ($r = 0.010$). This relationship strengthens in the five-year low-pass filtered time series ($r = 0.48$; Supplementary Fig. 3), indicating low-frequency co-variability between the two regions[35]. In CTL, Kuroshio and Gulf Stream SSTs exhibit significant positive correlations when the Gulf Stream leads by up to 4 years (Extended Data 1c), a lagged relationship reproduced in NAGA but absent in NPGA, further supporting an asymmetric influence from the Gulf Stream to the Kuroshio. In addition, the POGA experiment does not significantly reproduce CTL SST variability in either the Kuroshio or Gulf Stream region, nor does it capture the lagged correlation between the two regions.

A comparison of explained variance between CTL and the SST-restoring experiments further quantifies the modulatory contribution of Gulf Stream SST to Kuroshio SST. Quantitatively, North Atlantic SST in NAGA explains approximately 13% of the total variance in Kuroshio SST variability longer than interannual timescales, while tropical Pacific SST in POGA accounts for 4% (Fig. 1c). By contrast, the tropical Pacific forcing contributes to 17 % of the whole North Pacific mid-latitude more than the North Atlantic (10 %) (Fig. 1d). An independent statistical estimate yielded a consistent result (Supplementary Text 1), indicating a comparable contribution of Gulf Stream SST to Kuroshio SST.

Crucially, this asymmetric relationship substantially weakens in non-eddying ocean model experiments (NAGA-LR), despite the imposition of identical North Atlantic SST anomalies from CTL (Extended Data 1d-f). The correlation between Kuroshio SSTs in NAGA-LR and CTL is not statistically significant ($r = 0.07$), and the characteristic lagged correlation pattern, in which the Gulf Stream leads the Kuroshio, disappears. Accordingly, North Atlantic SST variability explains only 0.49% of Kuroshio SST variance in NAGA-LR. The NPGA-LR experiment also exhibits only a weak remote influence from the Kuroshio, consistent with NPGA.

Additional SVD and wavelet analyses provide complementary support for this interpretation (Supplementary Text 2). NAGA reproduces the coherent in-phase co-variability between Kuroshio and Gulf Stream SSTs and captures the low-frequency variability of Kuroshio SSTs, whereas the corresponding relationships are not reproduced in NPGA. These complementary analyses further support the interpretation that Gulf Stream SST variability remotely modulates Kuroshio SST variability.

**Wind-driven Kuroshio Extension adjustment**

Having established the influence of North Atlantic SST anomalies on Kuroshio SST variability, we next examine the underlying physical mechanisms, focussing primarily on NAGA. Lag-composite analysis for positive Kuroshio SST events shows that positive SSH anomalies gradually accumulate along the Kuroshio Extension (Fig. 2;

Extended Data 2). These positive SSH anomalies accumulate offshore east of Japan over timescales of up to five years, which is consistent with the significant lagged correlation observed between Kuroshio and Gulf Stream SST (Extended Data 1c), and then induce locally anticyclonic circulation anomalies and thereby shift the Kuroshio Extension northward. The resulting oceanic circulation anomalies transport warm Kuroshio-origin waters northward and contribute positively to surface temperature anomalies in the Kuroshio-Oyashio mixed water region.

Ocean mixed-layer heat budget analysis (Methods) further indicates that horizontal advection plays a central role in Kuroshio SST variability (Fig. 2d-h). As SST anomalies develop, ocean dynamics and surface heat fluxes show competing contributions to Kuroshio SST variations. Ocean dynamical contributions, mainly explained by horizontal advection, contribute positively to the mixed-layer temperature tendency, whereas surface heat fluxes make a negative contribution. These results suggest that ocean dynamics strengthen SST anomalies, while surface heat fluxes dampen them. The mixed-layer heat budget is therefore consistent with the mechanism in which the northward shift of the Kuroshio Extension enhances the northward advection of warm waters.

These SSH anomalies can be traced to low-frequency westward-propagating signals in the North Pacific, observed in both CTL and NAGA (Extended Data 3). In these simulations, SSH anomalies originate in the central North Pacific and propagate westward, consistent with observational evidence and the theoretical behavior of wind-driven oceanic Rossby waves[39,40]. Indeed, SSH reconstruction from wind stress anomalies using a linear Rossby wave model (Methods) reproduces SSH variability particularly well in the Kuroshio Extension (Supplementary Fig. 7). Comparison between the simulated SSH anomalies and the Rossby model reconstruction also reveals regions where they are not fully consistent (Extended Data 4). This inconsistency may reflect additional processes not captured by the linear Rossby wave, including nonstationary relationship between the wind stress field and SSH anomalies, as suggested for the North Atlantic[41].

Although individual Rossby-wave events differ between CTL and NAGA, these differences likely reflect wind stress variability associated with atmospheric internal variability in CTL rather than a discrepancy in the Rossby-wave interpretation itself.

Nevertheless, the NAGA ensemble mean captures the spatiotemporal patterns of SST and SSH variability in the Kuroshio Extension observed in CTL (Supplementary Fig. 8). Specifically, SSH and SST anomalies in the Kuroshio Extension show a significant correlation between NAGA and CTL, whereas no such relationship is found in NAGA-LR. In addition, NAGA-LR fails to reproduce these westward-propagating signals. This failure likely reflects that atmospheric responses to the North Atlantic are too weak to generate Rossby waves over the North Pacific (as discussed in the next section). Also, the model's coarse resolution may prevent representation of the Kuroshio Extension jet and related dynamics[42–44].

Our findings are consistent with recent studies suggesting that the northward shift in the Kuroshio Extension, along with associated SST warming, results from wind-induced Rossby wave propagation[45,46]. While the eddy-permitting ocean model used here captures the essential dynamics, higher-resolution simulations would allow for more representation of fine-scale processes. Nevertheless, the agreement between NAGA and CTL provides evidence that Gulf Stream SST anomalies remotely modulate low-frequency variability of the Kuroshio Extension through North Pacific wind stress anomalies and oceanic Rossby wave adjustment.

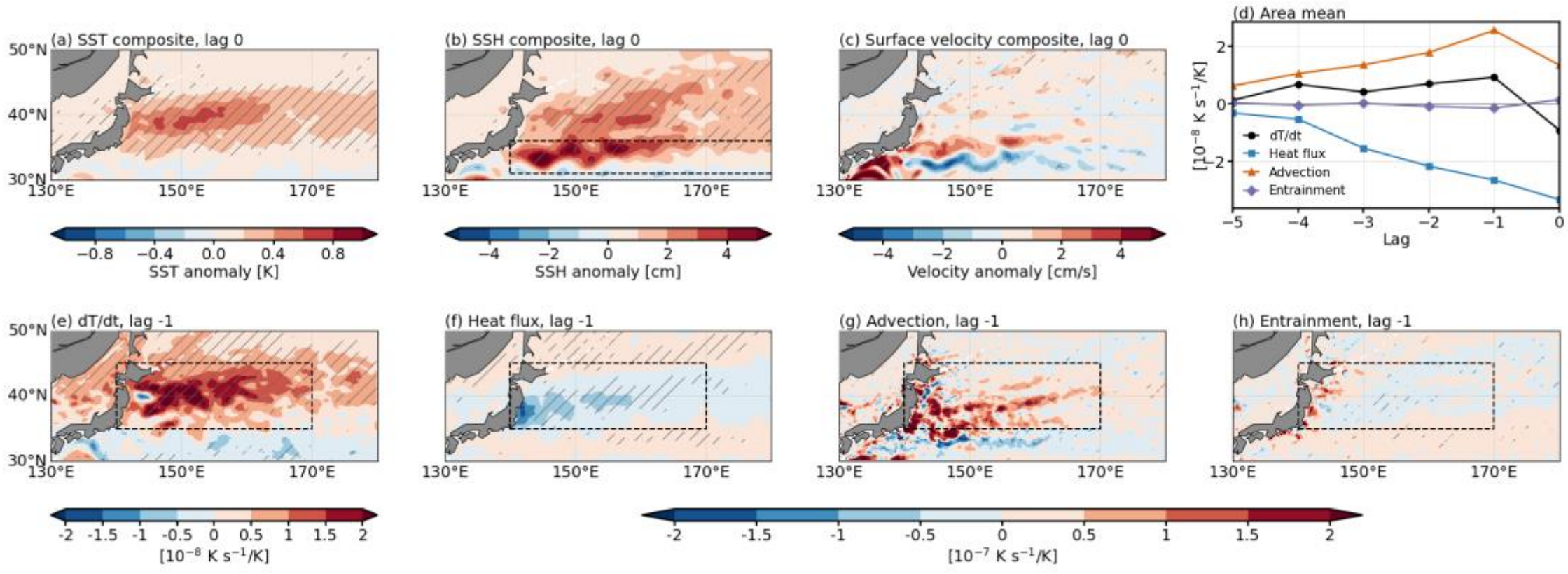

**Figure 2. Simulated surface ocean responses of the Kuroshio Extension associated with the SST warming.** (**a**-c) Composites of annual mean (a) SST, (b) SSH, and (c) surface velocity for annual mean Kuroshio SST exceeds +1 standard deviation for NAGA. Shading indicates the anomalies and hatching indicates statistically significant anomalies at the 90% confidence level. The black dashed area in (b) is area used in Extended Data 3. (d) Temporal evolution of annual mean mixed-layer temperature tendency, surface heat

flux contribution, and oceanic contribution (horizontal advection and entrainment) to the mixed-layer temperature tendency. All variables are area-averaged within black dashed boxes in (e)-(h) lag regression coefficients onto annual mean Kuroshio SST for NAGA ensemble means, and thus the unit is [$10^{-8}$K$s^{-1}$/K]. (e-h) As in (d), but for spatial patterns of each term regressed onto Kuroshio SST anomalies 1-year before the Kuroshio SST peaks. (e) Annual mean temperature tendency, (f) surface heat flux contribution, (g) horizontal advection, and (h) entrainment to the mixed-layer temperature tendency [$10^{-8}$K$s^{-1}$/K]. Hatchings indicate statistically significant regression coefficients. The black dashed box marks the Kuroshio region (140°E-170°E, 35°N-45°N).

**Influence of North Atlantic SST on Northern Hemisphere circulation anomalies**

Comparison of CTL and NAGA reveals that Kuroshio SST anomalies are primarily modulated by wind-driven SSH anomalies over the mid-latitude North Pacific induced by North Atlantic SST variability. This result indicates the presence of a hemispheric-scale atmospheric pathway originating from the North Atlantic. Previous studies have proposed both tropical and extratropical pathways through which basin-scale SST anomalies over the North Atlantic can influence the Pacific[29,30,34,47–51]. Therefore, we examine this atmospheric mechanism by first identifying the lower-tropospheric response and then evaluating how this response is associated with upper-level and hemispheric-scale circulation anomalies.

To evaluate the impact of Gulf Stream SST variability on large-scale atmospheric circulation, we regressed atmospheric circulation anomalies onto winter Gulf Stream SST anomalies in NAGA (Fig. 3a). Positive Gulf Stream SST anomalies are accompanied by a meridional dipole in SLP and lower-tropospheric circulation over the North Atlantic, particularly in early winter. This response is associated with a northward shift of the Gulf Stream SST front, which modifies lower-tropospheric baroclinicity and favours a northward displacement of the storm track (Supplementary Fig. 9). In addition to the imposed positive Gulf Stream SST anomalies, negative SST anomalies emerge in the subpolar gyre, where no SST restoring is applied. This meridionally contrasted SST pattern is consistent with the associated northward displacement of the storm track. A mixed-layer heat budget attributes the subpolar SST anomalies to anomalous surface heat fluxes and horizontal advection in early winter (Supplementary Fig. 10), in line with a previous high-resolution modelling study showing heat-flux-driven SST evolution[52].

The North Atlantic lower-tropospheric response further projects onto a hemispheric circulation pattern involving stratosphere–troposphere coupling. In NAGA, positive Gulf Stream SST anomalies are associated with positive geopotential height anomalies at 500hPa in the mid-latitudes and negative anomalies over the Arctic (Extended Data 5). In early winter, an NAO-like meridional dipole forms over the North Atlantic troposphere, accompanied by a strengthening of the stratospheric polar vortex. Subsequently, toward late winter, tropospheric anticyclonic anomalies develop over the North Pacific, leading to a weakening of the Aleutian Low.

Furthermore, these anomalies form a NAM-like structure that extends vertically from the stratosphere to the troposphere (Fig. 3b). The seasonal evolution of zonal-mean zonal wind anomalies further shows a downward extension of westerly anomalies from the strengthened polar vortex into the troposphere, providing a dynamical pathway through which Gulf Stream SST forcing modulates the North Pacific atmospheric circulation.

This interpretation distinguishes the Aleutian Low response in NAGA from the Pacific precursor-type pathway, in which basin-scale SST anomalies such as ENSO or PDO affect the Aleutian Low, the stratospheric polar vortex, and ultimately NAO-like anomalies[53–55]. Although NPGA exhibits an Aleutian Low response, it does not clearly extend a coherent stratospheric response or subsequent NAO-like anomalies over the North Atlantic (Extended Data 6, 7). This weaker response may reflect the limited spatial scale of the Kuroshio Extension SST forcing and its weaker anomalous meridional SST gradient compared with the Gulf Stream, highlighting the asymmetry between Gulf Stream and Kuroshio SST forcing.

AGCM sensitivity experiments (Methods) further support the role of North Atlantic SST forcing. When SST anomalies regressed onto Gulf Stream SSTs are prescribed over the North Atlantic, the AGCM reproduces the SLP anomalies seen in NAGA, including the weakening of the Aleutian Low during late winter (Supplementary Figure 15). The experiments also reproduce the strengthening of the polar vortex during early winter (Supplementary Text 4) and the downward-extending NAM-like zonal-wind anomalies from the stratosphere into the upper troposphere (Fig. 3c). These results demonstrate that the essential features of the hemispheric atmospheric response can be forced by Gulf Stream SST anomalies. By contrast, the AGCM experiment forced by the Kuroshio SST anomalies does not produce a clear stratosphere–troposphere coupled response (Extended Data 6 and 7), suggesting that the difference between NAGA and NPGA reflects the distinct atmospheric responses to the imposed SST forcing.

Finally, tropical-damping and low-resolution experiments further constrain the nature of this pathway. The co-variability between Gulf Stream and Kuroshio SSTs and the weakening of the Aleutian Low persist when tropical SST anomalies are suppressed (Supplementary Text 3), whereas the Aleutian Low response is not reproduced in NAGA-LR (Supplementary Text 5). These results indicate that the atmospheric pathway is rooted primarily in extratropical North Atlantic forcing and depends on resolving the Gulf Stream SST front, its meridional variability, and the associated atmospheric response.

NAGA, NPGA and AGCM experiments show that Gulf Stream forcing produces a pronounced stratosphere–troposphere coupled response. Observations and reanalysis provide consistent support for this atmospheric response (Supplementary Text 6). The atmospheric response to Gulf Stream SST can be interpreted within the framework of annular-mode dynamics proposed in previous studies, in which North Atlantic SST anomalies affect the polar vortex and subsequently influence tropospheric circulation[29,30,56–59], potentially through changes in upward planetary-wave propagation (Supplementary Text 4) associated with a shifted North Atlantic storm track[60]. Although the SST pattern examined in this study differs from basin-scale AMV and rather includes negative anomalies in the subpolar gyre, it can still be interpreted within the broader framework of annular-mode dynamics associated with North Atlantic SST forcing.

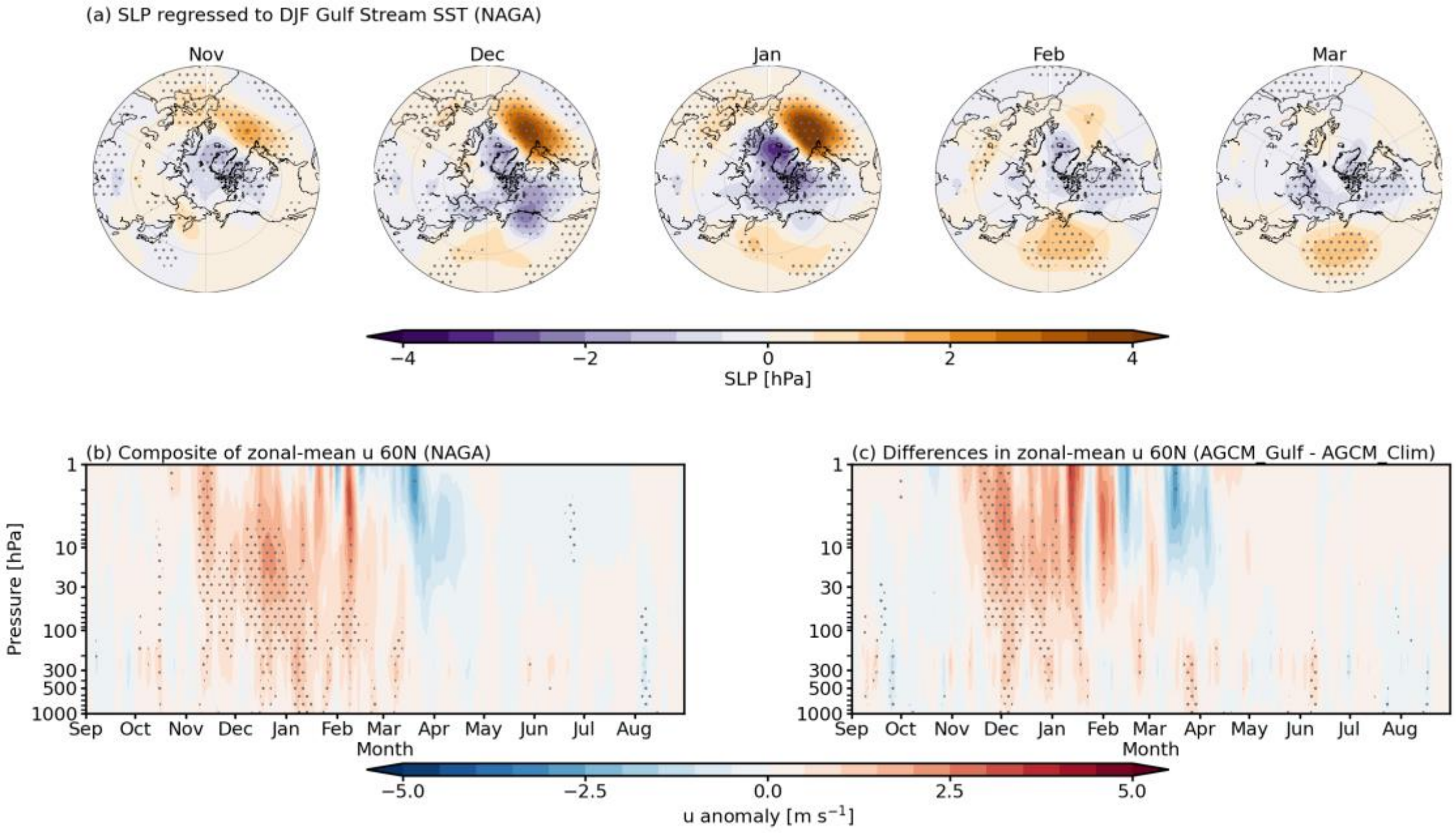

**Figure 3. Response of tropospheric and stratospheric circulation anomalies associated with Gulf Stream SST anomalies.** (a) Regression coefficients of monthly mean sea level pressure (SLP) [hPa/K] (shading) during boreal winter (from November to March) for the NAGA ensembles. SLPs are regressed to the DJF-mean SST anomalies in the Gulf Stream region. Stippling indicates regression coefficients statistically significant at the 90% confidence level. (b) Composites of zonal-mean zonal wind anomalies at 60°N when DJF Gulf Stream SST exceeds +1 standard deviation for the NAGA ensembles. Stippling indicates anomalies statistically significant at the 90% confidence level. (c) As in (b), but for differences between AGCM_Gulf and AGCM_Clim.

### Resolution dependence of the Gulf Stream to North Pacific pathway in CMIP6 models

To assess the robustness of resolution-dependent atmospheric responses to Gulf Stream SST variability, we performed a multi-model analysis using CMIP6 data. We examined the relationship between Gulf Stream SST and North Atlantic atmospheric circulation anomalies across multiple HighResMIP models[61,62] (Methods), supplemented by CTL and CTL-LR simulations. High-resolution models with eddy-permitting ocean components consistently reproduce a characteristic positive NAM-like response to positive Gulf Stream SST anomalies (Extended Data 8a). This response is characterized by a meridional dipole structure over the North Atlantic, negative anomalies over the Arctic, and positive anomalies over the North Pacific, consistent with reanalysis data. Conversely, low-resolution models with non-eddying ocean components tend not to capture this distinctive atmospheric pattern (Extended Data 8b).

A key difference between NAGA and NAGA-LR lies in the extent to which Gulf Stream SST anomalies induce a NAM-like atmospheric response and consequent Aleutian Low anomalies. This finding provides mechanistic insight into the resolution dependence of the Gulf Stream SST-NAM relationship observed across CMIP6 models. Specifically, high-resolution models reproduce the dipole-like SST anomalies in the North Atlantic (Supplementary Fig. 22), consistent with the mechanism discussed in the previous section and with a previous high-resolution modelling study[52]. Relationships among Gulf Stream SST, the NAO and the Aleutian Low further indicate that high-resolution models tend to reproduce the remote influence on the North Pacific more consistently (Extended Data 8c). Models with eddy-permitting ocean components are more likely to capture the NAO-like coupled atmosphere–ocean variability associated with Gulf Stream SST anomalies and, consequently, the anticyclonic circulation anomalies over the North Pacific.

Although our findings demonstrate the resolution-dependent coupling processes, coupled model control simulations alone are insufficient to clearly establish causality in the North Atlantic SST–NAM relationship. Under pre-industrial control or 1950-control conditions, SST variations can be passively triggered by internal atmospheric variability. This ambiguity likely accounts for the uncertain stratospheric associations to Gulf Stream SST variability across high-resolution models (Supplementary Fig. 23).

**Implications**

This study shows that Gulf Stream SST variability plays a modulatory role, accounting for approximately 13% of the low-frequency internal variability in Kuroshio SST. We identify a comprehensive dynamical mechanism wherein Gulf Stream SST anomalies induce NAM-like atmospheric responses, which leads to wind stress anomalies over the North Pacific. These, in turn, modulate Rossby wave adjustments in the Kuroshio Extension (Schematic figure in Fig. 4a). While previous studies have proposed potential links between the North Atlantic and the Kuroshio Extension, the present findings advance our understanding of Gulf Stream-Kuroshio interactions by identifying a specific stratospheric pathway. Importantly, this mechanism is well captured in climate models that incorporate high-resolution ocean components.

The proposed mechanism provides an explanation for recent unprecedented ocean warming across the mid-latitude of the North Pacific. The 2019–2020 North Atlantic SST dipole was accompanied by a positive NAO-like circulation pattern and North Pacific anticyclonic anomalies, followed by westward-propagating positive SSH anomalies that persisted in the western Kuroshio Extension from 2021 onward (Extended Data 9). This sequence likely contributed to the observed northward shift of the Kuroshio Extension[63] and East Asian heatwaves[64].

Since recent warming in the North Pacific has partly been linked to tropical Pacific forcing[65], we further estimated the contribution of the North Atlantic remote influence relative to tropical Pacific (Fig. 4b-g). These experiments include a set of SSP2-4.5 scenario simulations and additional sensitivity simulations in which observed SST anomalies are restored in the Gulf Stream region, the tropical Pacific, or both regions (Methods). The SSP2-4.5 scenario simulation does not reproduce the observed warming pattern. However, restoring observed Gulf Stream SST anomalies produces more pronounced SST anomalies primarily in the Kuroshio Extension, while tropical Pacific SST anomalies mainly affect the eastern North Pacific mid-latitude. When both observed SST forcings are included, the ensemble captures the observed basin-scale warming pattern (Fig. 4c, f). These results suggest that Gulf Stream SST anomalies contributed to the recent extreme warm SST anomalies, particularly over the western mid-latitude North Pacific. Moreover, given that this remote influence is associated with NAO-like atmospheric circulation anomalies, the latest decadal predictions indicating a tendency toward a positive NAO phase[66] are consistent with the possibility that high SST conditions in the North Pacific could persist over the next decade.

Such extreme warming events pose substantial risks to marine ecosystems and may have cascading impacts on fisheries, coastal communities, and broader socioeconomic systems[67]. It is therefore essential to assess whether the relationship between the Gulf Stream and the Kuroshio will intensify under continued global warming, as well as to evaluate the potential socioeconomic consequences. Moreover, these findings provide a new framework for understanding decadal climate variability[68,69] and highlight the crucial importance of high-resolution climate models for enhancing decadal prediction skill in the Northern Hemisphere, offering a potential pathway to address long-standing challenges such as the signal-to-noise paradox.

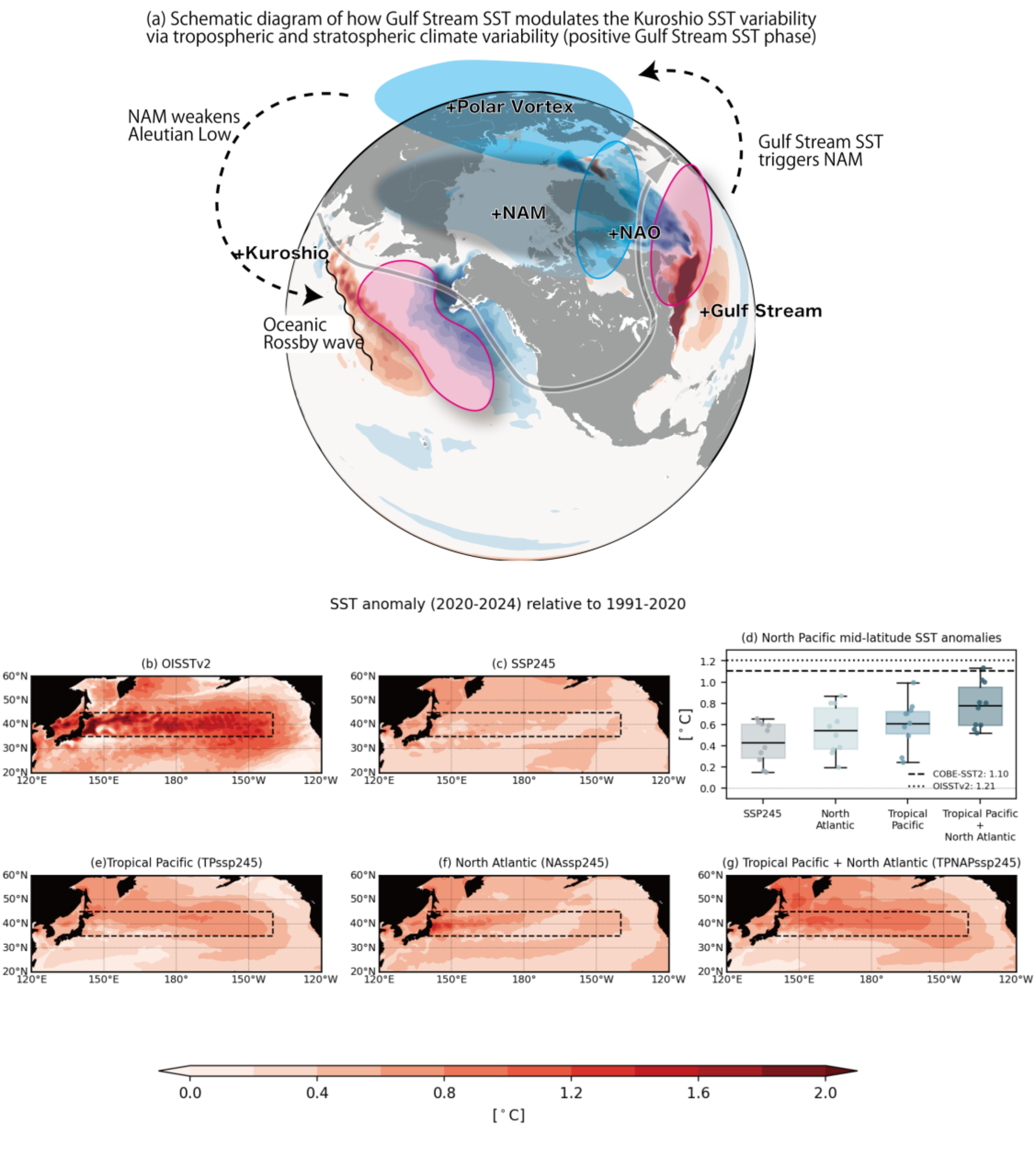

**Figure 4. Schematic diagram of the proposed mechanism linking Gulf Stream SST variability to Kuroshio SST variability and simulated relative contributions of external forcings and inter-basin forcings for North Pacific abnormal warming during 2020-2024.** (a) Schematic illustration of the influence of Gulf Stream SST

variability on Kuroshio SST via stratospheric and tropospheric pathways, as proposed in this study. Arrows indicate key processes, including NAM-like atmospheric responses, weakening of the Aleutian Low, and westward Rossby wave propagation. (b) observed (OISSTv2) and (c) simulated (MIROC6subhires SSP2-4.5) SST anomalies for 2020-2024 relative to 1991-2020 mean. (d) Regional-mean SST anomalies over the mid-latitude North Pacific (35°–45° N, 140° E–140° W) averaged for 2020–2024 relative to 1991–2020. Boxes show the interquartile range across ensemble members, black lines indicate ensemble means, whiskers span the full ensemble range, and points denote individual members; dashed and dotted horizontal lines show COBE-SST2 and OISSTv2 observations, respectively. As in (c) but for simulations for SSP2-4.5 with (e) tropical Pacific, (f) North Atlantic mid-latitude, (g) both tropical Pacific and North Atlantic mid-latitude forcings.

## Methods

### Observational and Reanalysis data

Monthly sea surface temperature (SST) data from the Centennial In Situ Observation-Based Estimates of the Variability of SST and Marine Meteorological Variables version 2 (COBE-SST2)[70] and the Hadley Centre Sea Ice and Sea Surface Temperature (HadISST)[71] were used for the period 1900–2024. National Oceanic and Atmospheric Administration (NOAA) Optimum Interpolation SST (OISST) version 2 High Resolution Dataset (OISSTv2)[72] are used for the period 1982-2024. Daily sea level anomaly (SLA) products from Copernicus Marine Service (CMEMS) (DUACS DT2014)[73] with a horizontal resolution of 1/4° × 1/4° were utilized for the period 1993–2023. Sea level pressure, geopotential height, and zonal wind data from ERA5[74] were used for the period 1950–2024.

### Definitions of climate indices

Gulf Stream and Kuroshio SST anomalies are defined as area-averaged SST anomalies over the North Atlantic (50°-80°W, 35°N-45°N) and North Pacific (140°E-170°E, 35°N-45°N)[35]. The North Atlantic Oscillation (NAO) index is also defined as the principal component of the leading EOF mode of the geopotential height anomalies at 500hPa over the North Atlantic domain (60°W-0°, 30°N-70°N).

**Statistical analysis**

For linear correlation and regressions analysis, a two-tailed t-test was applied at the 90% confidence level. The degrees of freedom were estimated following Bretherton et al.[75] The explained variance was calculated as the square of the correlation coefficient. Singular value decomposition (SVD) is applied to SST anomalies in the North Pacific (120°E-180°E, 25°N-55°N) and the North Atlantic (30°W-90°W, 25°N-55°N). Wavelet analysis[76] was applied to the SST timeseries.

**Climate model simulations**

We used the sixth version of the Model for Interdisciplinary Research on Climate (MIROC6)[37]. The atmospheric component employs a T85 spectral truncation, corresponding to a horizontal resolution of approximately 1.4° for both latitude and longitude. It includes 81 vertical levels, with the model top at 0.004 hPa. The ocean component employs a tripolar coordinate system, with a longitudinal grid spacing of 1° and meridional grid spacing varying from 0.5° near the equator to 1° in the mid-latitudes, and 62 hybrid σ-z vertical levels. The ocean component uses the surface mixed layer parameterization[77] and the eddy isopycnal diffusion parameterization[78]. A high-resolution version of MIROC6, i.e., MIROC6subhires[35,79], was also used. The atmospheric component is identical to that of MIROC6, but the ocean horizontal resolution is increased to 0.25° × 0.25° with the same number of vertical levels. The surface mixed layer parameterization is the same as that in MIROC6, but the eddy isopycnal diffusion is not applied between the equator and 60°N/ 60°S. After reaching quasi-equilibrium in the spin-up phase, an additional 200-year (1000-year) integration was performed for MIROC6subhires (MIROC6[80]) using a preindustrial external forcing in accordance with the CMIP6 protocol[38]. Hereafter, we refer the last 100 years of output from MIROC6subhries (MIROC6) as CTL (CTL-LR) (Supplementary Table 1).

**SST-restoring experiments**

To highlight the relative role of SST in the Kuroshio and Gulf Stream, we conducted ensemble SST-restoring (i.e., pacemaker) experiments using MIROC6subhires, in which SST was restored toward a target field. In these experiments, an additional heat flux was added to the surface heat flux at each atmosphere–ocean coupling time step to relax the simulated SST toward a target SST field derived from CTL. The target SST field was constructed from the monthly mean SST of CTL, interpolated in time between adjacent months, and remapped onto the AGCM grid prior to the restoring calculation. Therefore, SST restoring primarily constrains basin-scale SST variability and associated meridional displacement of SST fronts rather than individual oceanic mesoscale eddies.

The SST restoring was applied over the North Atlantic (30°–75° W, 30°–50° N) in the North Atlantic-Global Atmosphere (NAGA)[81] experiment, over the North Pacific (130°–180° E, 30°–50° N) in the North Pacific-Global Atmosphere (NPGA) experiment, and over the Tropical Pacific in the Pacific Ocean-Global Atmosphere (POGA) experiment[15,65] (Supplementary Fig. 1).

The nudging flux added to the surface heat flux is calculated as:

$$Nudging\ flux = \frac{\rho C_p h}{\tau} \times \left(SST_{\text{target}} - SST_{model}\right). \qquad (1)$$

$\rho$ (=1027 kg m$^{-3}$) is the density of the seawater, $C_p$(=4187 J kg$^{-1}$ K$^{-1}$) is the specific heat of seawater, $\tau$ (= 15days) is the restoring time scale, and h (=50 m) is assumed to be surface mixed layer affected by SST restoring. Here, $SST_{\text{target}}$ denotes the time-interpolated monthly mean SST from CTL on the AGCM grid, and $SST_{model}$ is the model SST used at each atmosphere–ocean coupling step.

Outside the restoring regions, the restored SST anomaly was tapered linearly to zero within 6° from the restoring boundaries (Supplementary Fig. 1). Initial conditions are taken from CTL, spaced by more than 10 years interval. Both NAGA and NPGA were integrated for 100 years with 10 ensemble members under the preindustrial conditions as CTL. When SST variability is constrained in the Kuroshio or Gulf Stream regions, correlations between the area-averaged SSTs from the ensemble mean and CTL exceed 0.94 (Supplementary Fig. 2). This indicates that SST-restoring works well as expected.

Analogous SST-restoring experiments were also conducted for the North Atlantic and North Pacific using MIROC6, referred to here as NAGA-LR and NPGA-LR, respectively. In these experiments, initial conditions were taken from the corresponding low-resolution control experiment (CTL-LR), with initial conditions spaced by more than 10 years, whereas the target SST field used for restoring was derived from the monthly mean SST of CTL. This configuration was adopted so that the imposed SST target is identical between the high- and low-resolution restoring experiments.

The ensemble mean is interpreted as the model-estimated response common to the imposed SST anomalies, while deviations of individual members include internally variability and modulation of internal variability by the SST forcing. For each SST-restoring experiment, the explained variance of the Kuroshio SST index was estimated as the squared correlation between CTL and the ensemble mean. Spreads were calculated as the standard deviation of the corresponding values across individual members.

**Scenario experiments**

To assess the relative contributions of observed SST anomalies in the Gulf Stream and tropical Pacific to the recently observed anomalous Kuroshio SST conditions, we conducted additional 10-member ensemble pacemaker experiments with MIROC6subhires under the SSP2-4.5 scenario[82] for 2015–2024. Initial conditions were taken from the MIROC6subhires historical ensemble simulations. The standard 10-member SSP2-4.5 ensemble simulations were compared with three corresponding pacemaker ensembles, in which SST was restored over the North Atlantic (NAssp245), the tropical Pacific (TPssp245), and both regions simultaneously (TPNAssp245). The SST-restoring regions were the same as those used in NAGA and POGA (Supplementary Fig. 1). The target SST field was constructed by adding observed SST anomalies from COBE-SST2, defined relative to its 1970–2014 climatology, to the 1970–2014 climatological SST in the MIROC6subhires historical simulations. Otherwise, the restoring procedure, including the restoring timescale and additional heat-flux formulation, was identical to that described above.

**Atmospheric model sensitivity experiment**

Atmospheric responses to SST anomalies in the Gulf Stream and Kuroshio regions were investigated using the atmospheric component of MIROC6 and MIROC6subhires (Supplementary Table 3). In the control simulation (AGCM_Clim), the atmospheric model was forced with the monthly climatology of SST and sea ice concentration derived from CTL. Additional experiments were conducted in which constant SST anomalies are superimposed over the North Atlantic (30°–75° W, 30°–50° N; AGCM_Gulf) and the North Pacific (130°–180° E, 30°–50° N; AGCM_Kuro), with the anomalies linearly tapered to zero within 6° outside these regions. The imposed anomalies were defined as the linear regression patterns associated with the area-averaged Gulf Stream SST timeseries (50°-80°W, 35°N-45°N) and Kuroshio SST (140°E-170°E, 35°N-45°N) time series in CTL. Differences from AGCM_Clim represents the atmospheric responses to SST anomalies in the Gulf Stream and Kuroshio regions, respectively. All simulations were integrated for 21 years with 10 ensemble members initialized from CTL at intervals 10 years. The first year was excluded as spin-up, and the remaining 20 years from all ensemble members were combined, yielding 200 samples for comparison.

**CMIP6 simulations**

CMIP6 multi-model ensemble data were analysed to explore the relationship between North Atlantic SST and Northern Hemisphere atmospheric circulation. Control-1950 experiments from HighResMIP[61] were used to exclude the influence of external forcing. Models with both non-eddying and eddy-permitting ocean components were selected: CNRM-CM6-1[83], CNRM-CM6-1-HR[84], EC-Earth3P[85], EC-Earth3P-HR[86], ECMWF-IFS-LR[87], ECMWF-IFS-MR[88], HadGEM3-GC31-LL[89], and HadGEM3-GC31-MM[90] (Supplementary Table. 4). A 100-year segment (1950-2049) of each simulation was analysed.

**Baroclinic linear Rossby wave model**

To investigate the role of the oceanic Rossby wave on SSH anomalies in the Kuroshio Extension regions, we implemented the baroclinic linear Rossby wave model[36,40,63]. When the 1.5-layer reduced gravity ocean is assumed, the wind-forced linear vorticity equation is expressed below under the long-wave assumption;

$$\frac{\partial \eta}{\partial t} - c_R \frac{\partial \eta}{\partial x} = -\frac{g'}{g} w_{\mathrm{Ek}}. \quad (2)$$

Here, $\eta$ is the SSH anomaly, $c_R$ is the phase speed of first baroclinic Rossby waves, $g$ is the gravity acceleration, $g'$ is the reduced gravity, and $w_{\mathrm{Ek}} (= \boldsymbol{k} \cdot \nabla \times \left(\frac{\tau}{\rho_0 f}\right))$ is the Ekman pumping velocity anomalies. f is the Coriolis parameter, $\rho_0$ is the reference density, **k** is the vertical unit vector, and $\tau$ is the wind stress vector. Following the previous studies[36,40,46], we adopt $c_R = 4.2$ [cm/s] and $g' = 0.03$ [m/s$^2$]. Integrating

Equation (2) along the characteristic line yields the analytical solution for;

$$\eta(x,y,t) = \eta[x + c_R(t - t_0), y, t_0] - \frac{g'}{g}\int_{t_0}^{t} w_{\mathrm{Ek}}\,[x + c_R(t - t'), y, t']\,dt'. \quad (3)$$

The first term on the right hand side represents westward propagation of SSH anomalies, while the second term indicates integrated Ekman pumping anomalies along westward wave propagations.

**Ocean mixed layer heat budget**

The temperature balance within the ocean mixed layer was evaluated following previous studies[79,91,92]:

$$\frac{\partial T_m}{\partial t} = \frac{Q_{\mathrm{net}} - q_d}{\rho C_p H} - \mathbf{u}_m \cdot \nabla T_m - \frac{\Delta T}{H} w_e + \mathrm{res}\,. \quad (4)$$

Here, $T_m$ denotes the mixed layer temperature. The first term on the right-hand side represents the contribution from surface heat flux, where $Q_{net}$ is the net surface heat flux, $q_d$ is the downward solar insolation penetrating through the bottom of the mixed layer, $\rho$ (=1027 kg m$^{-3}$) is the density of seawater, $C_p$(=4187 J kg$^{-1}$ K$^{-1}$) is the specific heat of seawater, and $H$ is the mixed layer depth. $H$ was defined as the depth at which the density exceeds the surface density by 0.125 kg m$^{-3}$. The second term represents horizontal advection, and the third term is the entrainment through the bottom of the mixed layer. $\Delta T$ (=$T_m - T_{-H-10m}$) is the temperature difference ($T_{-H-10m}$ is the temperature 10 m below the mixed layer bottom). $w_e$ is the entrainment velocity based on the following formulation:

$$w_e = \frac{\partial H}{\partial t} + \mathbf{u}_{-h} \cdot \nabla H + w_{-h}. \quad (5)$$

$\mathbf{u}_{-h}$ and $w_{-h}$ are horizontal and vertical velocity at the mixed layer bottom. When the right-hand-side of Eq. (5) is negative, $w_e$ is set to 0. The last term of Eq. (4) is a residual term consisting of diffusion, detrainment and other nonlinear processes.

**Storm track activity, Eady growth rate, and Eliassen-Palm flux**

The tropospheric atmospheric responses to SST anomalies were analyzed by diagnoses of storm track activity and Eady growth rate. Storm track activity is defined as meridional eddy heat flux at 850hPa (i.e.,$\overline{v'T'}$ ), where $v'$ and $T'$ represented 8-day high-pass filtered daily meridional wind and temperature. The overbar indicates the monthly mean since we mainly focus on the monthly mean of the storm track activity in this study.

To measure the lower tropospheric baroclinicity, we introduced the Eady growth rate (EGR)[93] at the 700hPa;

$$\mathrm{EGR} = 0.31 f \left| \frac{\partial y}{\partial z} \right| N^{-1}. \quad (6)$$

The planetary wave propagation and its influence on the acceleration/deacceleration of zonal-mean zonal wind was also diagnosed using Eliassen-Palm flux (E-P flux). The E-P flux was formulated in pressure coordinate on the sphere following the previous study[94];

$$\mathbf{F} = \left\{ F_{(\phi)}, F_{(p)} \right\} = a \cos \phi \left\{ -\overline{u'v'} + \psi \frac{\partial \bar{u}}{\partial p}, -\overline{u'\omega'} - \psi \left[ \frac{\partial (\bar{u} \cos \phi)}{a \cos \phi \, \partial \phi} - f \right] \right\}, \quad (7)$$

where

$$\psi = \frac{\overline{v'\theta'}}{\partial \bar{\theta} / \partial p} = - \frac{\overline{v'T'}}{\kappa \bar{T}/p - \partial \bar{T} / \partial p}. \quad (8)$$

Here, $\phi$ represents latitude, p pressure, $\omega$ the velocity in longitude, latitude, pressure coordinates, $\theta$ the potential temperature, T the temperature, a the earth's radius, f the Coriolis parameter and $\kappa$ the ratio of the gas constant to the specific heat at the constant pressure. Overbars indicate zonal average at constant $\phi$ and p, and primes are anomalies from the zonal means. The contribution of wave activity to the acceleration and deacceleration of $\bar{u}$ (i.e., $\frac{\partial \bar{u}}{\partial t}$) can be calculated by the divergence of scaled by $\frac{1}{a \cos \phi}$:

$$\frac{1}{a \cos \phi} \nabla \cdot \mathbf{F} = \frac{1}{a \cos \phi} \left( \frac{\partial (F_{(\phi)} \cos \phi)}{a \cos \phi \, \partial \phi} + \frac{\partial F_{(p)}}{\partial p} \right). \quad (9)$$

**Data availability**

The HadISST dataset was downloaded from the Met Office website (https://www.metoffice.gov.uk/hadobs/hadisst/). The COBE-SST2 dataset was downloaded from the NOAA Physical Sciences Laboratory website (https://psl.noaa.gov/data/gridded/data.cobe2.html). OISST v2 is available at https://www.esrl.noaa.gov/psd/data/gridded/data.noaa.oisst.v2.highres.html. The altimeter products were produced and distributed by Copernicus Marine Environment Monitoring Service (http://marine.copernicus.eu). ERA5 data were downloaded from ECMWF (https://doi.org/10.24381/cds.f17050d7). The CMIP6 experimental dataset was downloaded from the ESGF website (https://esgf-node.llnl.gov/projects/cmip6/). The Model outputs generated in this study are available from the corresponding author upon reasonable request.

## Code availability

The code used to generate the man figures in this study will be deposited in a public repository. The model code is available under restricted access in accordance with the developers' policy. Access can be obtained upon reasonable request by contacting the corresponding author.

**Acknowledgements**

We thank Hisashi Nakamura and Bo Qiu for their helpful comments. During the preparation of this article, the authors used GPT-5.5 for English editing. The authors then reviewed and edited the content as needed, and take full responsibility for the content of the final publication. The authors thank FORTE Science Communications (https://www.forte-science.co.jp/) for English language editing. This work was supported by MEXT program for the advanced studies of climate change projection (SENTAN) Grant Number JPMXD0722680395 (Y.Y., H.T., T.K., S.K., S.O.), JSPS KAKENHI Grant Number JP20H05729 (Y.Y.), JP22K14098 (Y.Y.), JP22H04487 (Y.Y., T.K., S.K.), JP23H01241 (T.K.), and JP23K13169 (T.K.), and JP24H00280 (Y.Y., H.T., S.K.). The model simulations were performed using Earth Simulator at the Japan Agency for Marine-Earth Science and Technology (JAMSTEC), Japan.

**Author contributions**

Y.Y. conceived this study, conducted simulations, performed the analysis, and wrote the draft. H.T. lead the development of the MIROC6 and MIROC6subhires. Y.K. and Y.Y. conducted the historical and SSP2-4.5 scenario experiments. Y.Y., H.T., T.K., S.K., S.O., and Y.K. discussed the results and contributed to the manuscript.

**Competing interests**

The authors declare no competing interests.

**Supplementary information**

Supplementary Information is available for this paper.

**Extended Data**

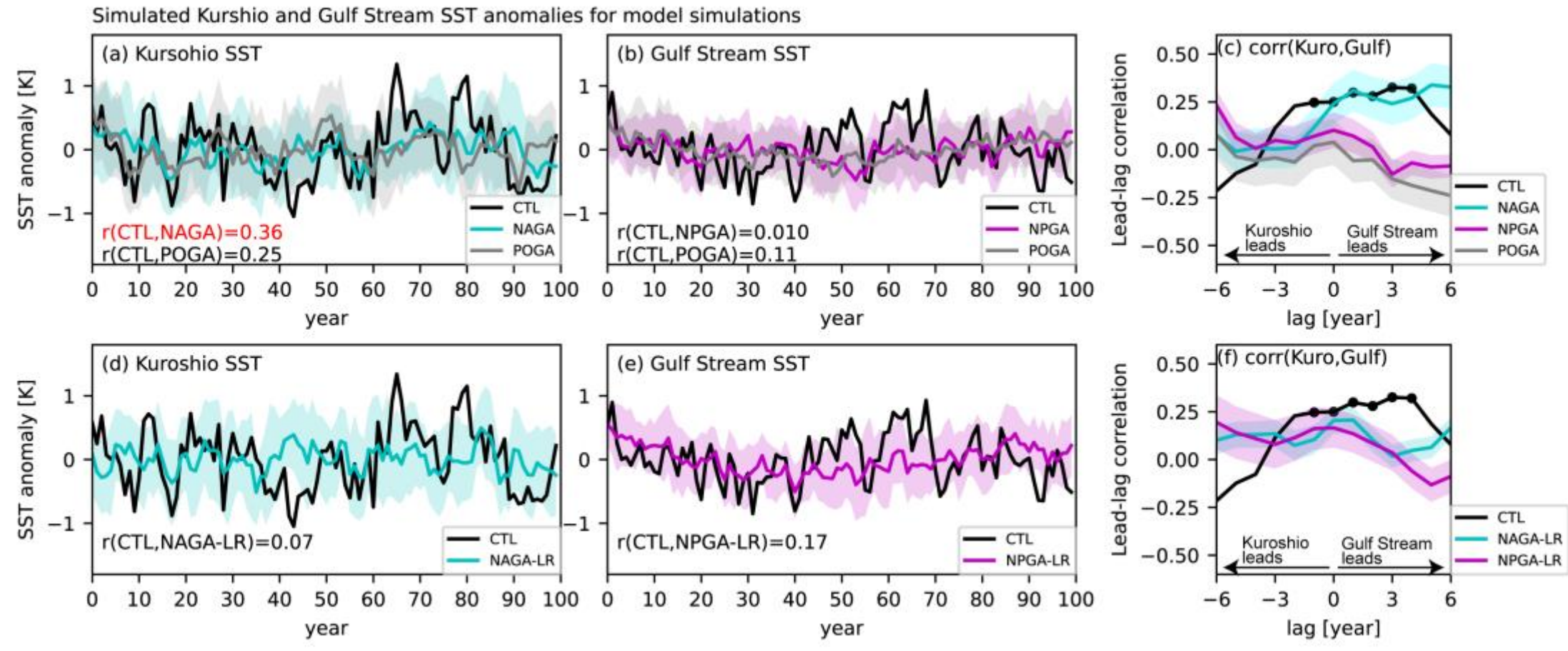

**Extended Data 1. Simulated impacts of Kuroshio and Gulf Stream SSTs on each other.** (a) Simulated annual mean SST anomalies in the Kuroshio region (140°E–170°E, 35°N–45°N), based on CTL (black), the ensemble mean of NAGA (cyan), and the ensemble mean of POGA (gray). Shading indicates ± 1 standard deviation across ensemble members. Correlation coefficient of NAGA and POGA to CTL is shown in the upper right; red text denotes statistical significance at the 90% confidence level, while a black correlation coefficient indicates no significance at the 90% level. (b) As in (a), but for SST anomalies in the Gulf Stream region (50°–80°W, 35°–45°N) from CTL, NPGA (magenta), and POGA. (c) Lag correlation between annual mean Kuroshio and Gulf Stream SST anomalies for CTL (black), NPGA (magenta), NAGA (cyan), and POGA (gray). Positive lags indicate that Gulf Stream SST leads. Solid black dots denote significant correlations in CTL. Shading indicates ±1 standard deviation across ensemble members. (d)–(f) As in (a)-(c), but for NPGA-LR and NAGA-LR.

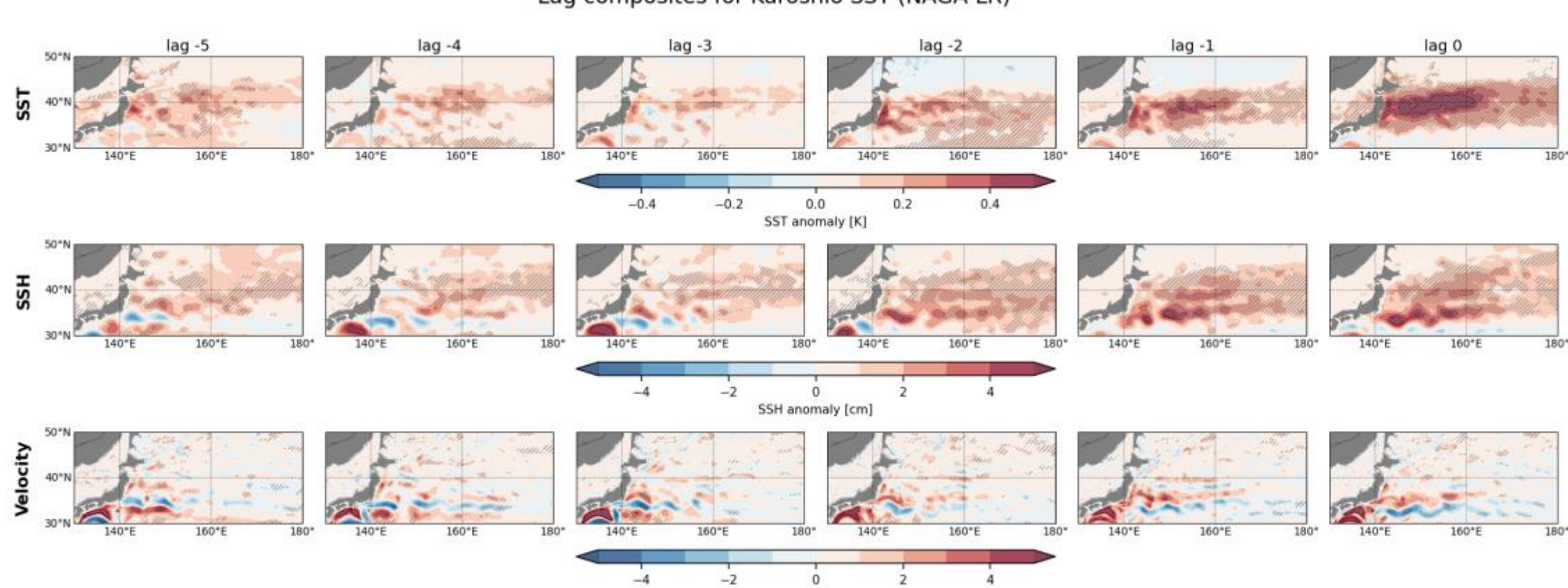

**Extended Data 2. Response of the Kuroshio Extension to positive SST anomaly events.** Lag composites of annual mean SST, SSH, and surface velocity when Kuroshio SST exceeds +1 standard deviation for NAGA ensemble means. Hatching indicates statistically significant anomalies at the 90% confidence level.

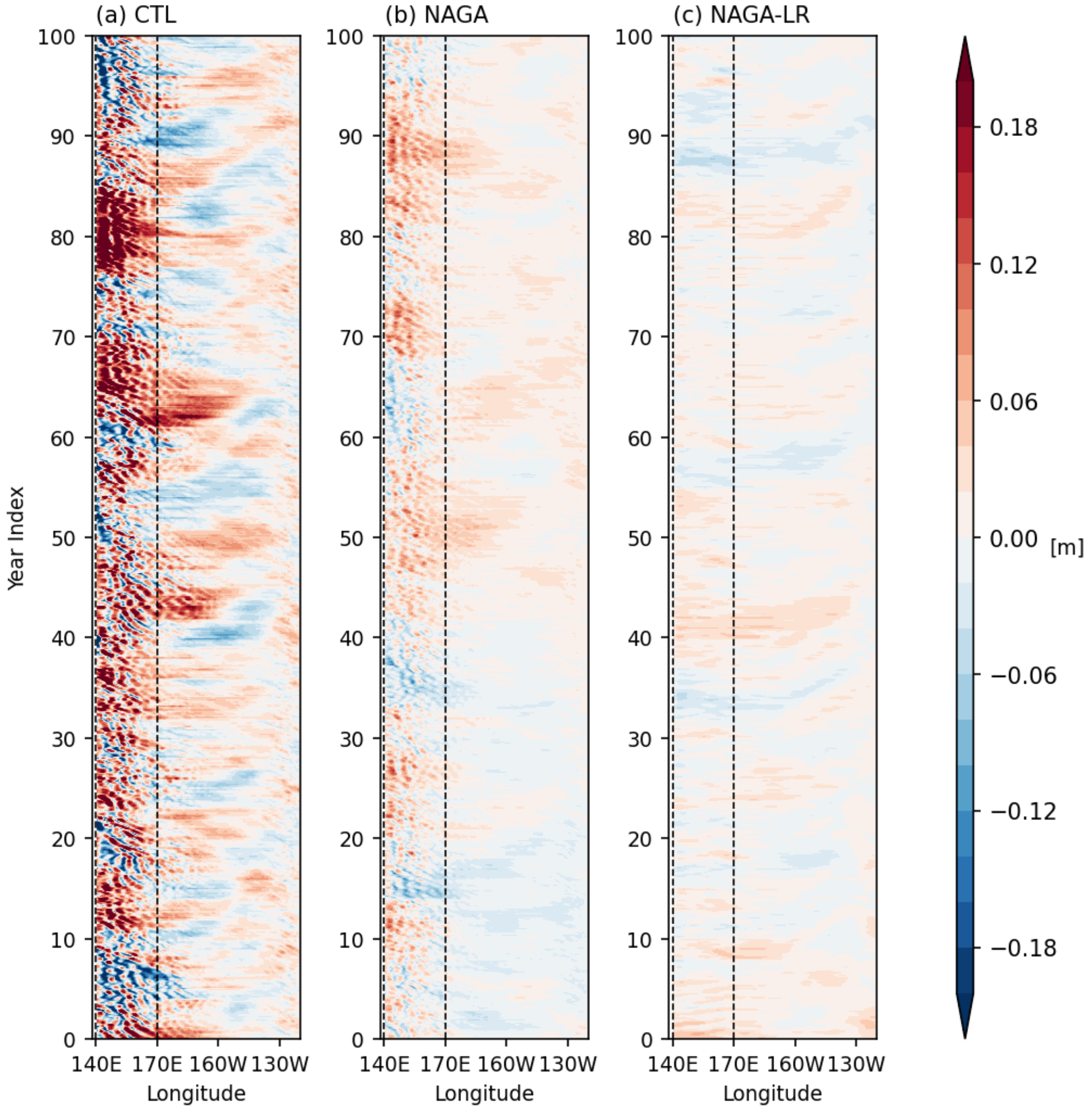

**Extended Data 3. Comparison of westward-propagating signals of SSH anomaly signals.** Monthly SSH anomalies [m] for **(a)** CTL, **(b)** NAGA, and **(c)** NAGA-LR, averaged over 31°N–36°N (black dashed box in Fig. 2b). Horizontal axis indicates longitude, while the vertical axis means year index in the model simulations. Dashed lines indicate the Kuroshio region as defined in this study (140°E–170°E).

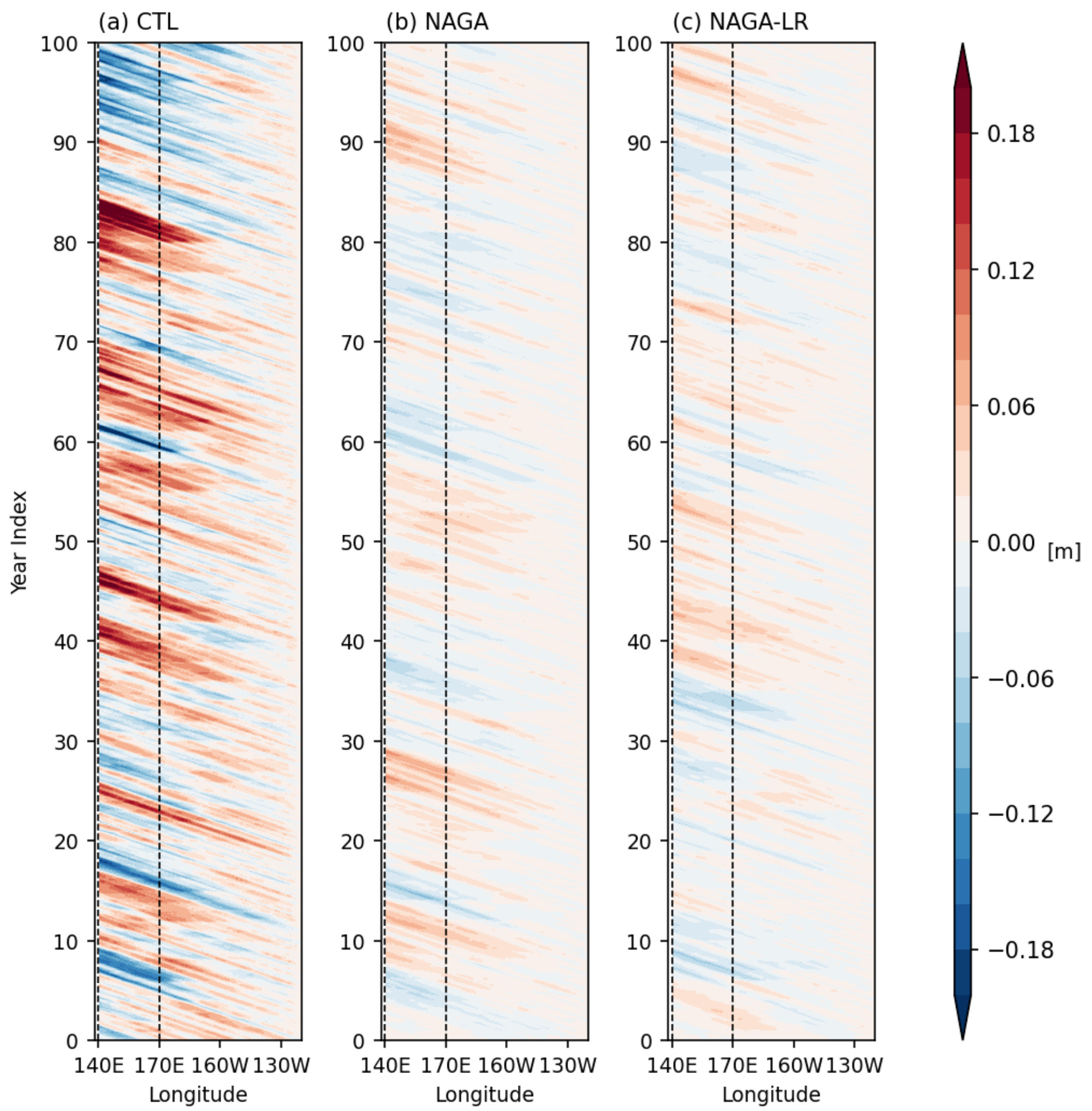

**Extended Data 4. SSH anomalies reconstructed by the linear Rossby wave model.** Monthly SSH anomalies [m] reconstructed from wind stress curl anomalies for (a) CTL, (b) NAGA, and (c) NAGA-LR, averaged over 31°N–36°N (Fig. 2b). Horizontal axis indicates longitude, while the vertical axis means year index in the model simulations. Dashed lines indicate the Kuroshio region as defined in this study (140°E–170°E).

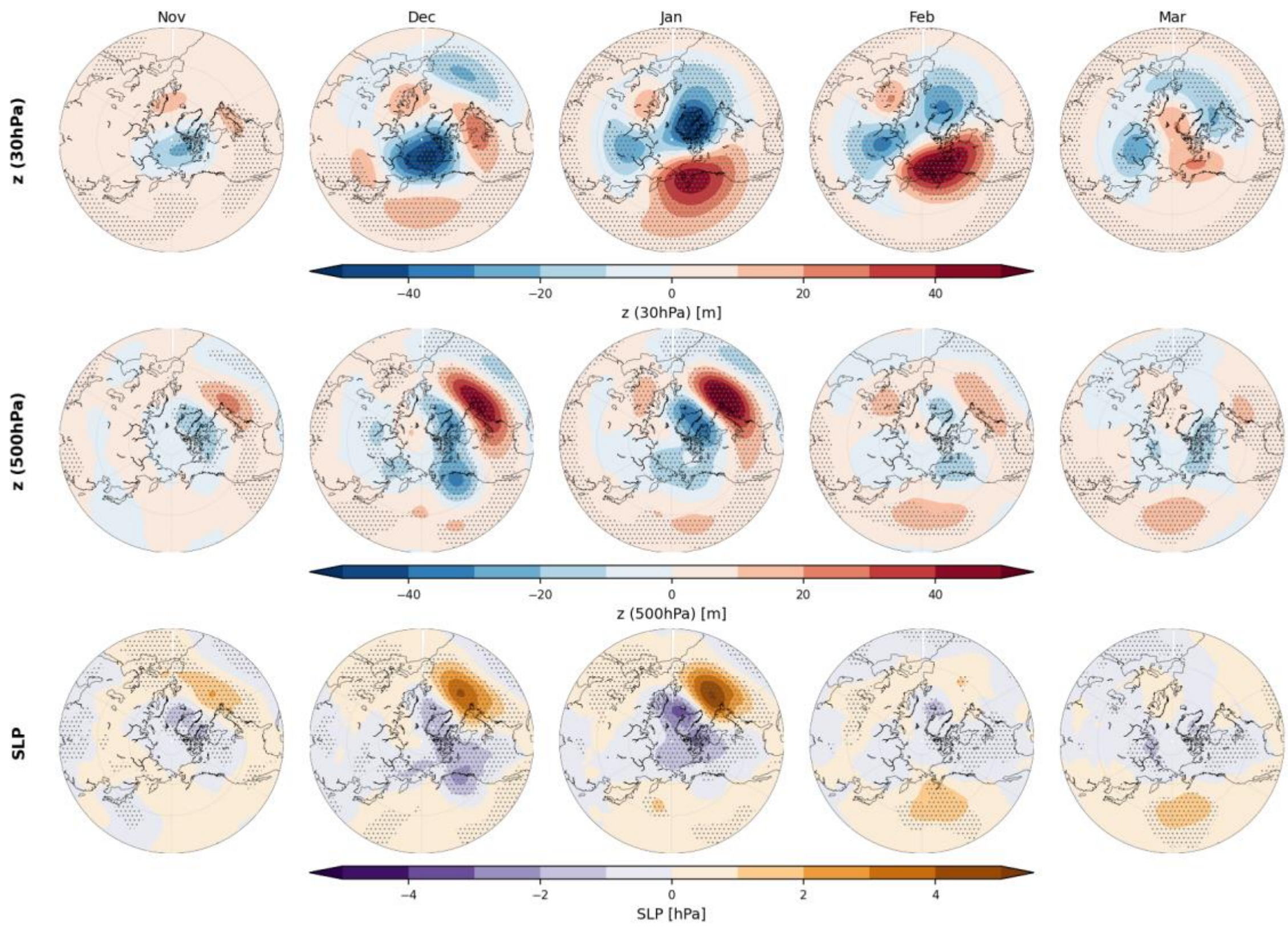

**Extended Data 5 Response of tropospheric and stratospheric circulation anomalies associated with Gulf Stream SST anomalies.** As in Fig. 3(a), but for regression coefficients of monthly mean geopotential height at 30 hPa (top), at 500hPa (middle), and sea level pressure (SLP; bottom) during boreal winter (from November to March) for the NAGA ensembles. All variables are regressed to the DJF-mean SST anomalies in the Gulf Stream region. Stippling indicates regression coefficients statistically significant at the 90% confidence level.

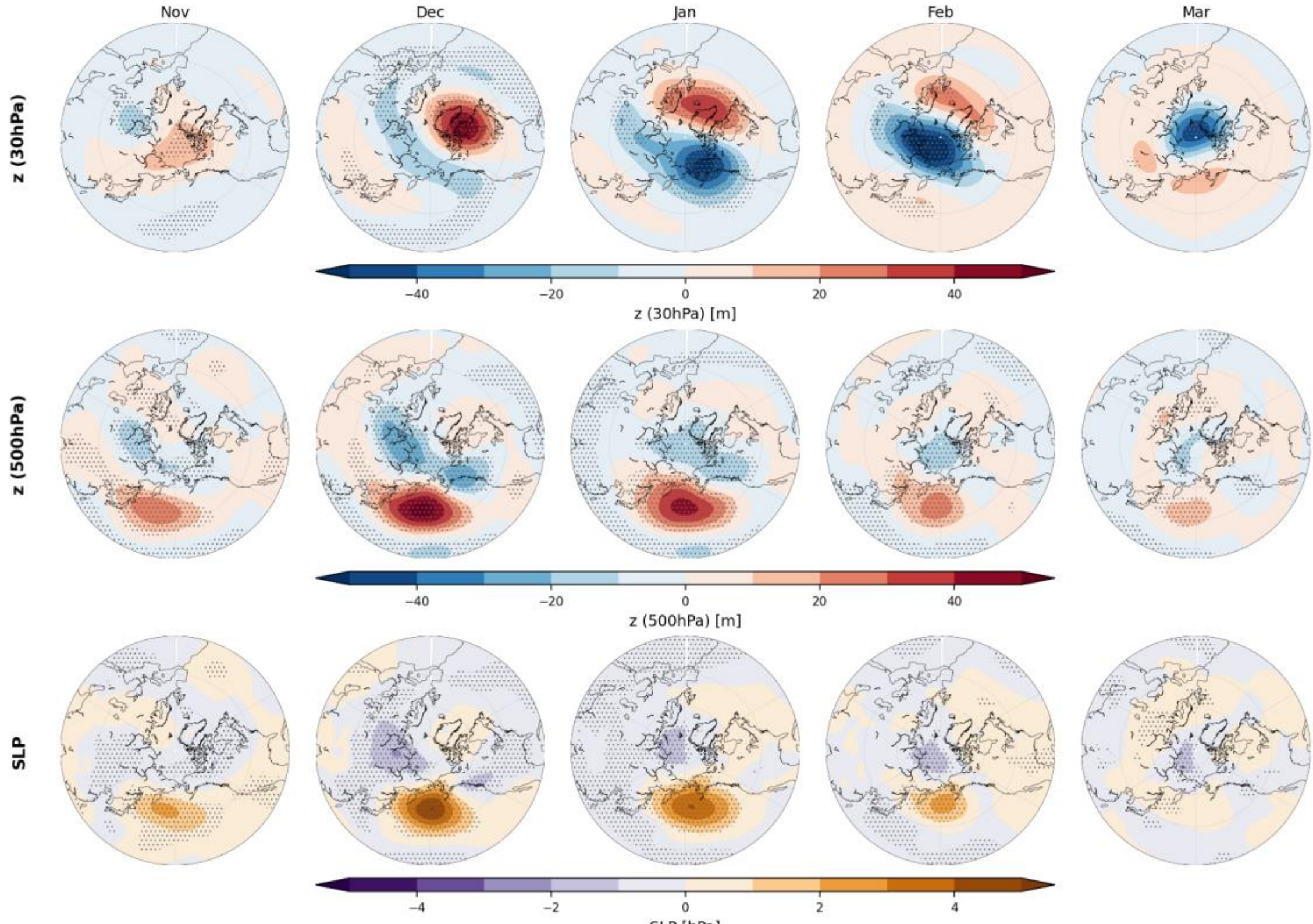

**Extended Data 6. Response of tropospheric and stratospheric circulation anomalies associated with Kuroshio SST anomalies.** As in Extended Data 5, but for regressions to Kuroshio SST anomalies for the NPGA ensembles.

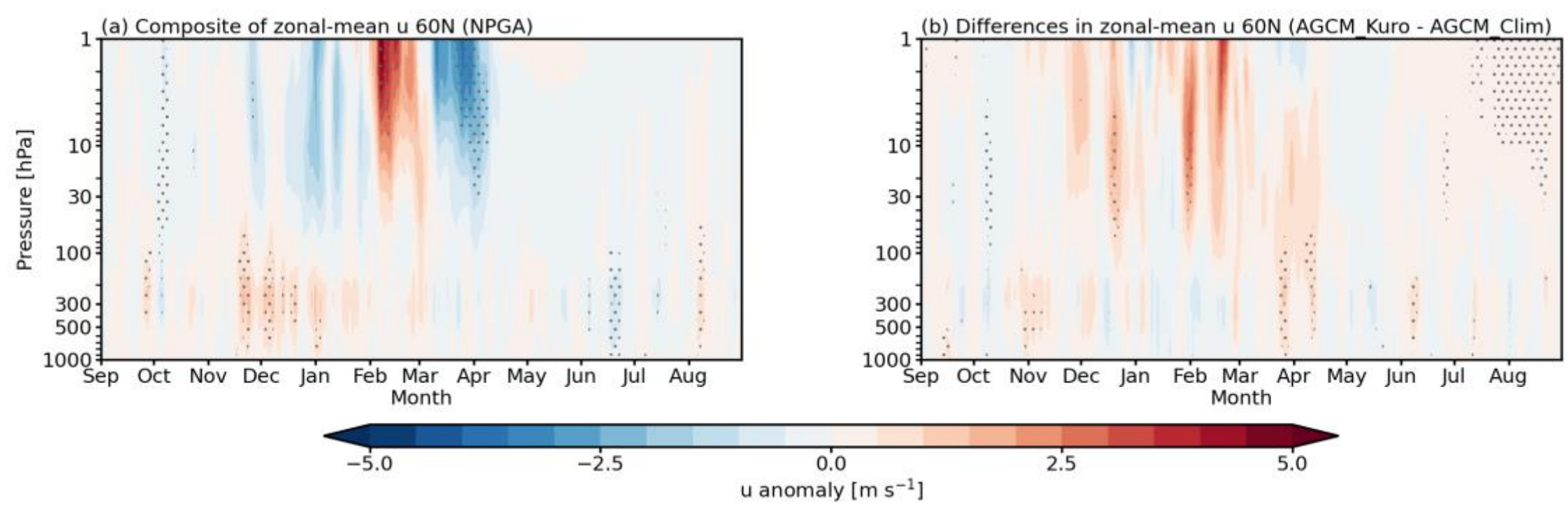

**Extended Data 7. Response of tropospheric and stratospheric circulation anomalies associated with Kuroshio SST anomalies.** As in Fig. 3 (b) and (c), but for responses associated with Kuroshio SST. (a) Composites of zonal-mean zonal wind anomalies at 60°N when DJF Kuroshio SST exceeds +1 standard deviation for the NPGA ensembles. Stippling indicates regression coefficients statistically significant at the 90% confidence level. (b) As in (a), but for differences between AGCM_Kuro and AGCM_Clim.

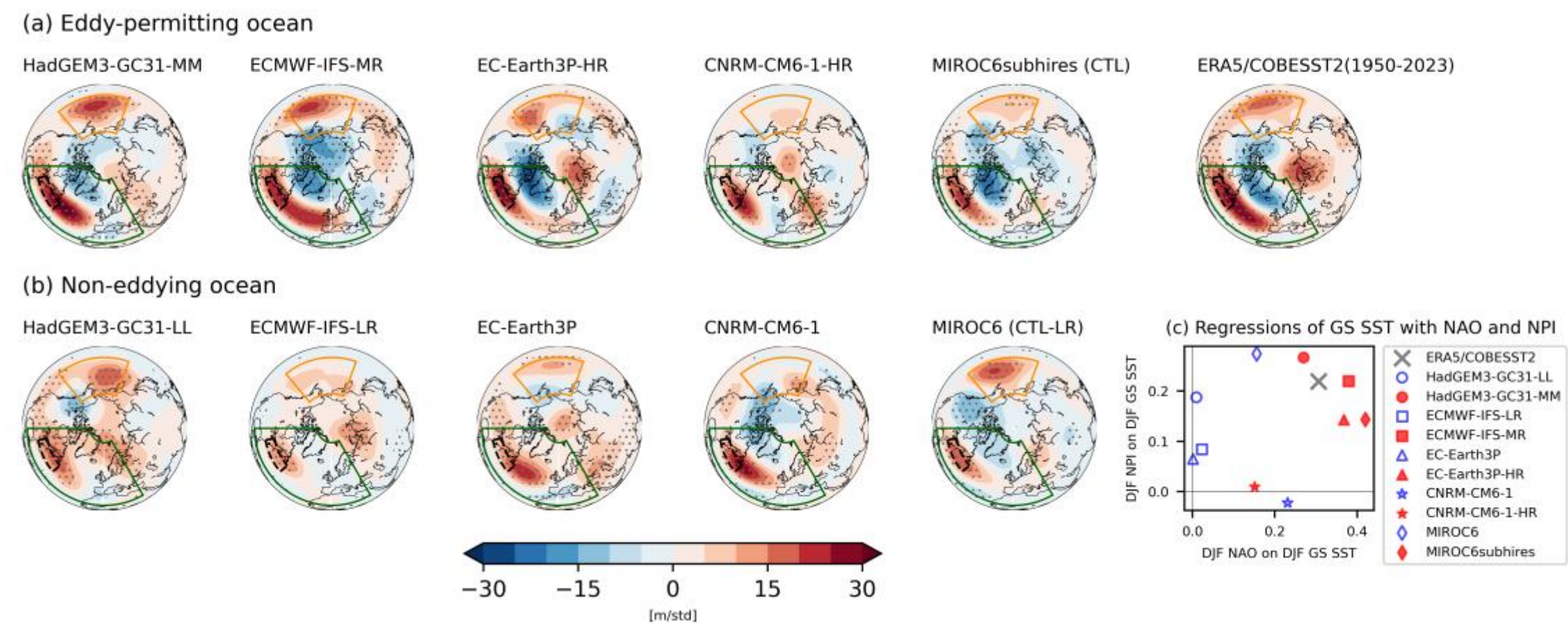

**Extended Data 8. Resolution dependence of the tropospheric circulation response to Gulf Stream SST anomalies in CMIP6 models.** DJF (December–January–February) mean geopotential height at 500 hPa [m] regressed onto normalized DJF mean SST anomalies in the Gulf Stream region (black dashed box) for CMIP6 models with **(a)** eddy-permitting and **(b)** non-eddying ocean model resolution (Methods). Stippling indicates regression coefficients statistically significant at the 90% confidence level. Observational geopotential height data are from ERA5, and SST data are from COBE-SST2 for the period 1950–2023. Both datasets are detrended prior to analysis. (**c)** Scatter plot showing regression coefficients of DJF NAO index (horizontal axis) and North Pacific index (NPI; Methods) (vertical axis) onto DJF Gulf Stream SST (GS SST) for each CMIP6 model. NAO index (NPI) was calculated from anomalies over the green (orange) areas in (a) and (b). Red (blue) symbols indicate eddy-permitting (non-eddying) ocean resolution. Since all indices are normalized, both regression coefficients are dimensionless.

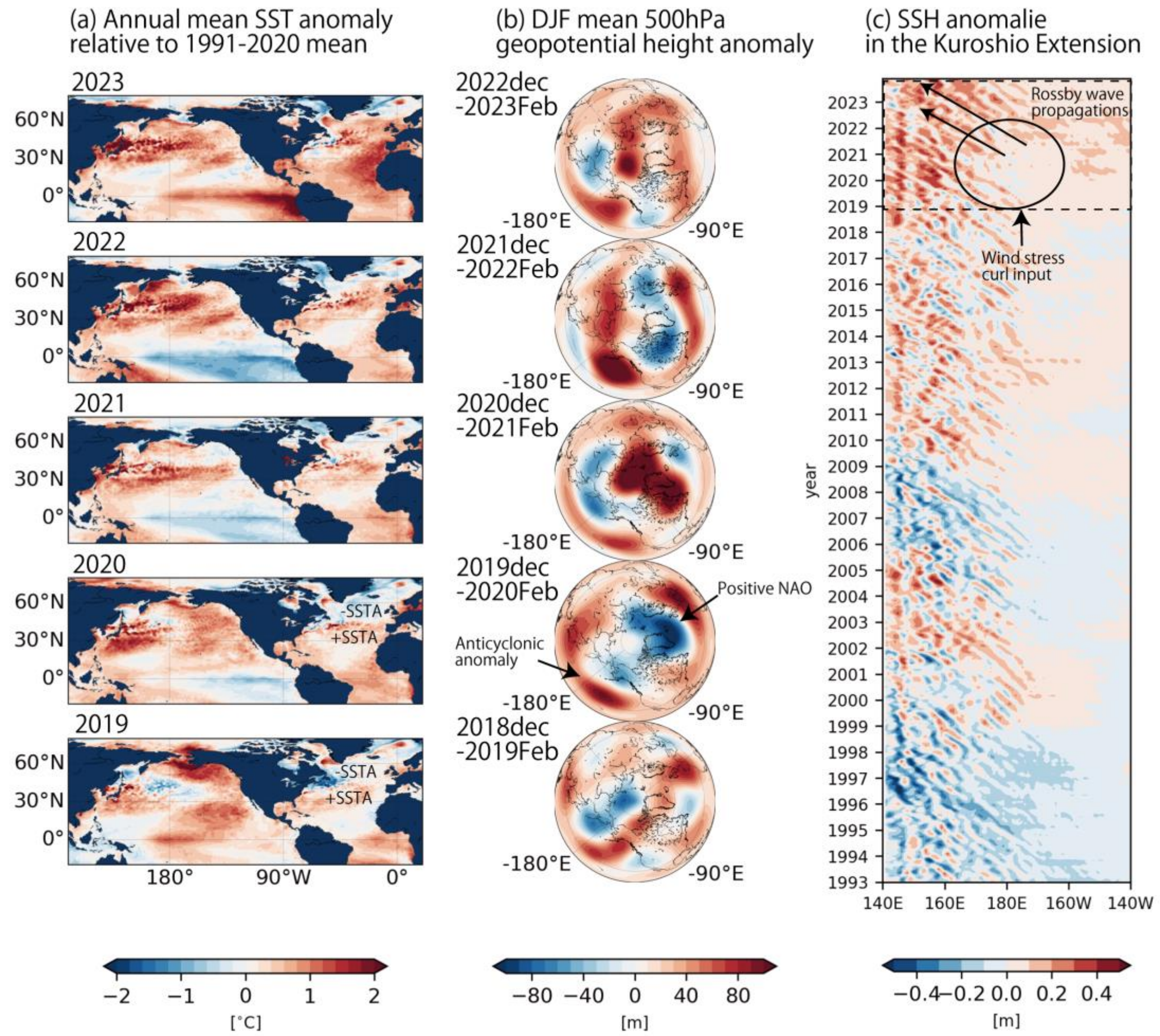

**Extended Data 9. Observational evidence for the recent SST, atmospheric circulation, and SSH anomalies in the Northern Hemisphere. (a)** Annual mean SST anomalies relative to the 1991–2020 climatology, based on OISSTv2 data for the period 2019-2023. **(b)** December–January–February (DJF) mean geopotential height anomalies at 500 hPa from ERA5, defined as deviations from the 1991-2020 climatology. Each panel is labelled by the corresponding DJF season (e.g., “Dec 2018 – Feb 2019”). **(c)** Observed sea surface height (SSH) anomalies in the Kuroshio Extension (32°N–36°N), based on CMEMS data. Monthly SSH anomalies are calculated as deviations from the 1993–2020 monthly climatology. "Year" on the y-axis denotes the January 1 for each year.

**Supplementary Data for**

**Gulf Stream contribution to recent North Pacific warming**

**Yoko Yamagami[1], Hiroaki Tatebe[1], Tsubasa Kohyama[2], Shoichiro Kido[1], Satoru Okajima[3], Yoshiki Komuro[1]**

[1]Japan Agency for Marine-Earth Science and Technology, Yokohama, Japan

[2]Department of Information Sciences, Ochanomizu University, Tokyo, Japan

[3]Institute of Life and Environmental Sciences, University of Tsukuba, Tsukuba, Japan

Corresponding author: Yoko Yamagami (y.yamagami@jamstec.go.jp)

†Research Center for Environmental Modeling and Application, Japan Agency for Marine-Earth Science and Technology, 3173-25 Showamachi, Kanazawaku, Yokohama, Kanagawa 236-0001, Japan

**Supplementary Table 1.** Summary of numerical experiments using coupled models. “Free” for the Gulf Stream, Kuroshio, and Tropics indicates that SST evolves freely without any SST restoring. “CTL” for the Gulf Stream, Kuroshio, and Tropics indicates that SST anomalies are restored to the monthly SST anomalies in CTL. The SST-nudging regions for each experiment are shown in Supplementary Figure 1.

| Name | Ocean nominal resolution | SST-restoring for CTL | | | Ensemble member | Period [year] |
|---|---|---|---|---|---|---|
| | | Gulf Stream | Kuroshio | Tropics | | |
| CTL | 0.25° | Free | Free | Free | 1 | 100 |
| NAGA | 0.25° | CTL | Free | Free | 10 | 100 |
| NPGA | 0.25° | Free | CTL | Free | 10 | 100 |
| CTL-LR | 1.0° | Free | Free | Free | 1 | 100 |
| NAGA-LR | 1.0° | CTL | Free | Free | 10 | 100 |
| NPGA-LR | 1.0° | Free | CTL | Free | 10 | 100 |
| POGA | 0.25° | Free | Free | CTL | 10 | 100 |

**Supplementary Table 2.** As in Supplementary Table 1, but for tropical SST-damping experiments. "Clim" in the Tropics column indicates that equatorial SSTs are nudged to climatological SST from CTL. The SST-restoring regions for each experiment are shown in Supplementary Figure 11.

| Name | Ocean nominal resolution | SST-restoring | | | Ensemble member | Period [year] |
|---|---|---|---|---|---|---|
| | | Gulf Stream | Kuroshio | Tropics | | |
| CTL-NoATL | 0.25° | Free | Free | Clim (Atlantic) | 1 | 100 |
| CTL-NoINPAC | 0.25° | Free | Free | Clim (Indo-Pacific) | 1 | 100 |
| CTL-NoTRO | 0.25° | Free | Free | Clim (whole tropics) | 1 | 100 |

**Supplementary Table 3.** Summary of AGCM sensitivity experiments.

| Name | Prescribed SST anomalies | | | Ensemble member | Simulation years |
|---|---|---|---|---|---|
| | Gulf Stream | Kuroshio | Tropics | | |
| AGCM_ Clim | No | No | No | 10 | 21 |
| AGCM_ Gulf_SST | Yes | No | No | 10 | 21 |
| AGCM_ Kuro_SST | No | Yes | No | 10 | 21 |

**Supplementary Table 4.** HighResMIP models analysed in this study, along with their nominal ocean horizontal resolution (km), and variant ID.

| model | Ocean nominal resolution [km] | Variant-ID |
|---|---|---|
| CNRM-CM6-1 | 100 | r1i1p1f2 |
| CNRM-CM6-1-HR | 25 | r1i1p1f2 |
| EC-Earth3P | 100 | r1p1i1f2 |
| EC-Earth3P-HR | 25 | r1p1i1f2 |
| ECMWF-IFS-LR | 100 | r1p1i1f1 |
| ECMWF-IFS-MR | 25 | r1p1i1f1 |
| HadGEM3-GC31-LL | 100 | r1p1i1f1 |
| HadGEM3-GC31-MM | 25 | r1p1i1f1 |

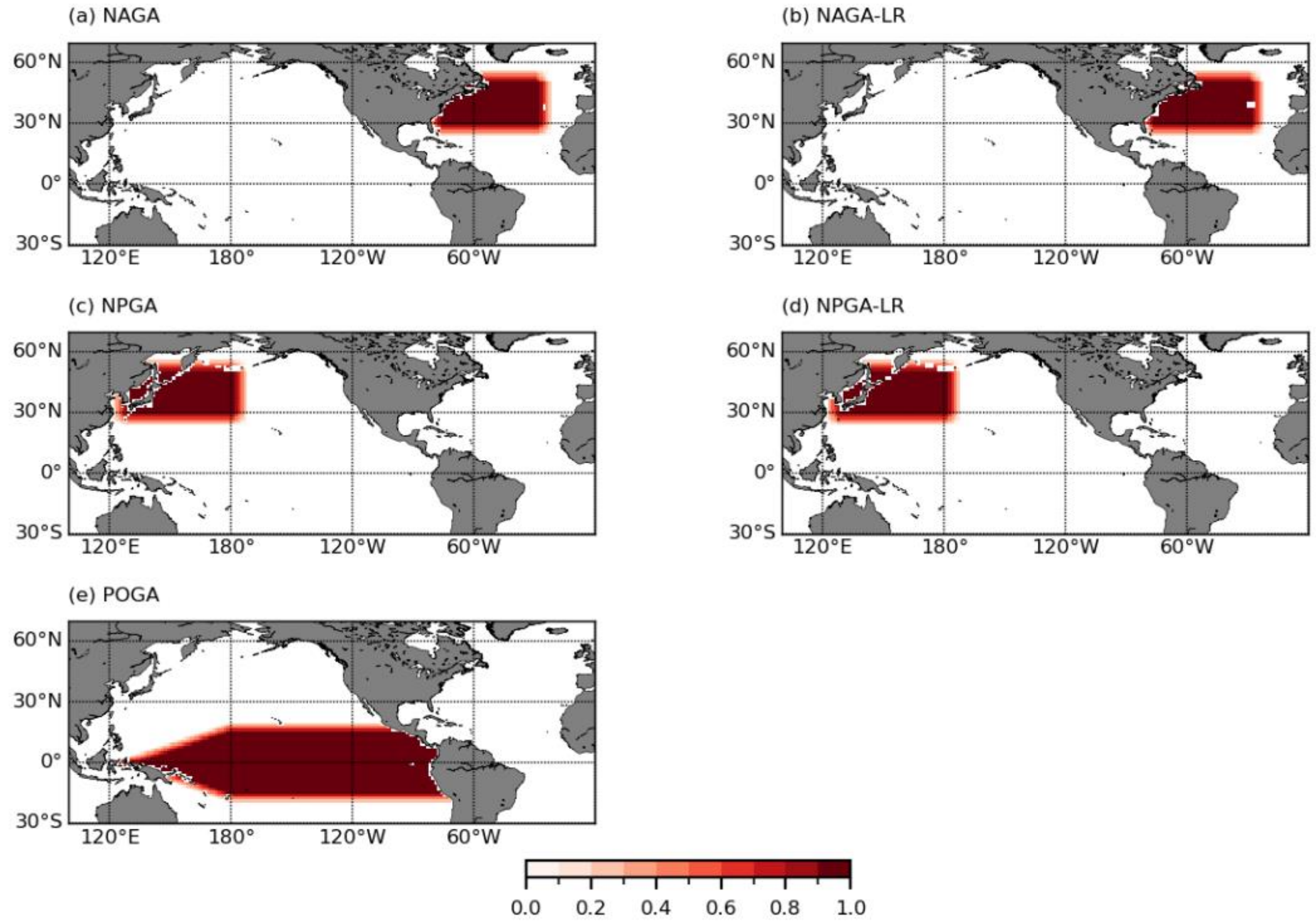

**Supplementary Figure 1. SST-restoring regions for each experiment. (a)** Normalized weighting factors applied to the nudging heat flux in NAGA. The restored SST anomalies taper linearly to zero within 6° of the nudging region boundary. (**b-d)** As in **(a)**, but for **(b)** NAGA-LR, **(c)** NPGA, **(d)** NPGA-LR, and (e) POGA.

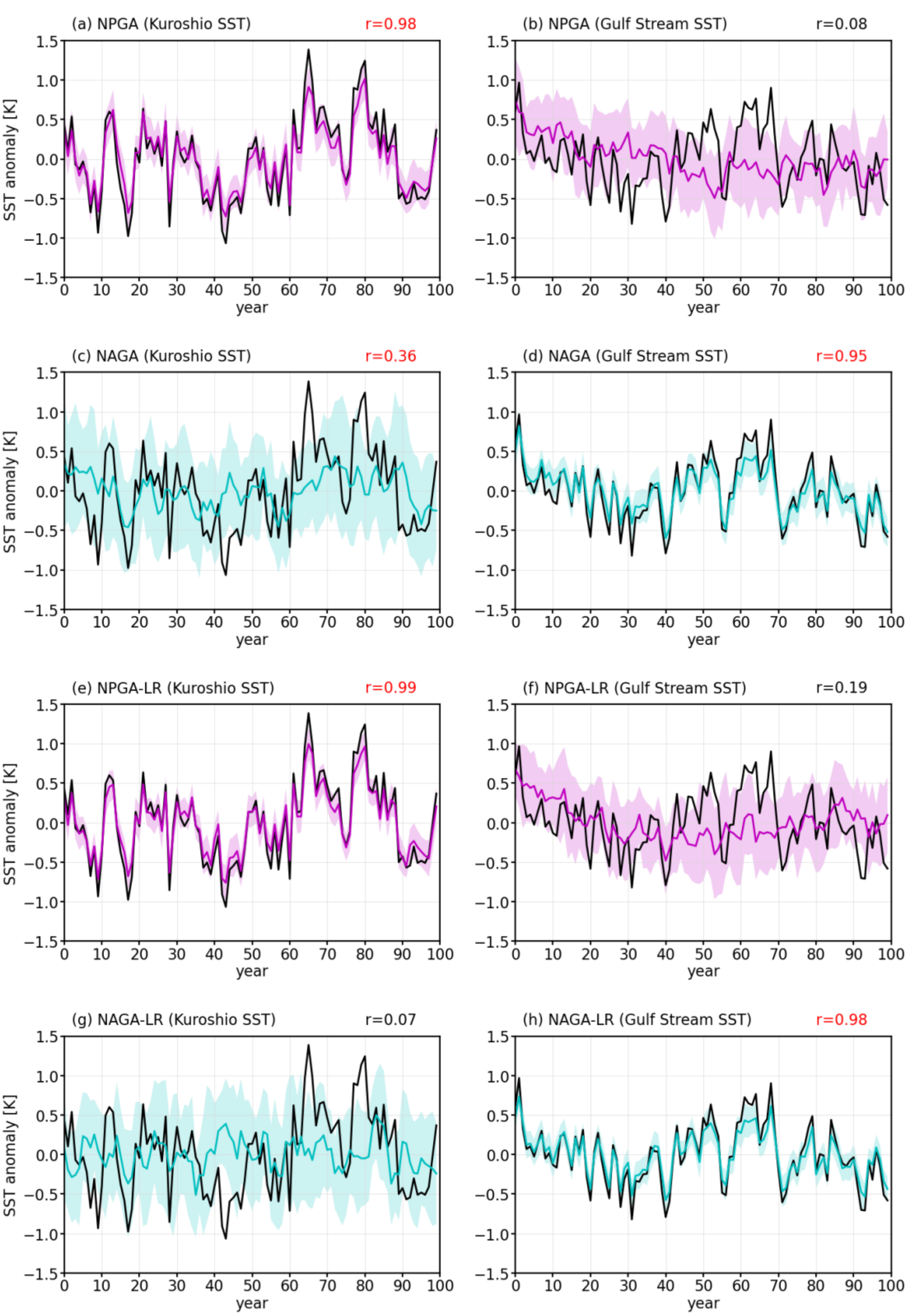

**Supplementary Figure 2. Simulated time series of annual mean SST anomalies for the Kuroshio and Gulf Stream regions.** (a) Annual mean SST anomalies [°C] averaged

over the Kuroshio region (140°E–170°E, 35°N–45°N; see boxed area in Fig. 1a) for CTL (black) and the ensemble mean of NPGA (magenta). Shading denotes one standard deviation across ensemble members. The correlation coefficient between CTL and NPGA ensemble mean is shown at the top right, red indicates statistical significance at the 90% confidence level. In contrast to Fig. 1, the linear trend is retained in each time series. **(b)** As in **(a)**, but for SST anomalies averaged over the Gulf Stream region (50°–80°W, 35°–45°N; black box in Fig. 1a). **(c–h)** As in **(a)** and **(b)**, but for **(c, d)** NAGA, **(e, f)** NPGA-LR, and **(g, h)** NAGA-LR.

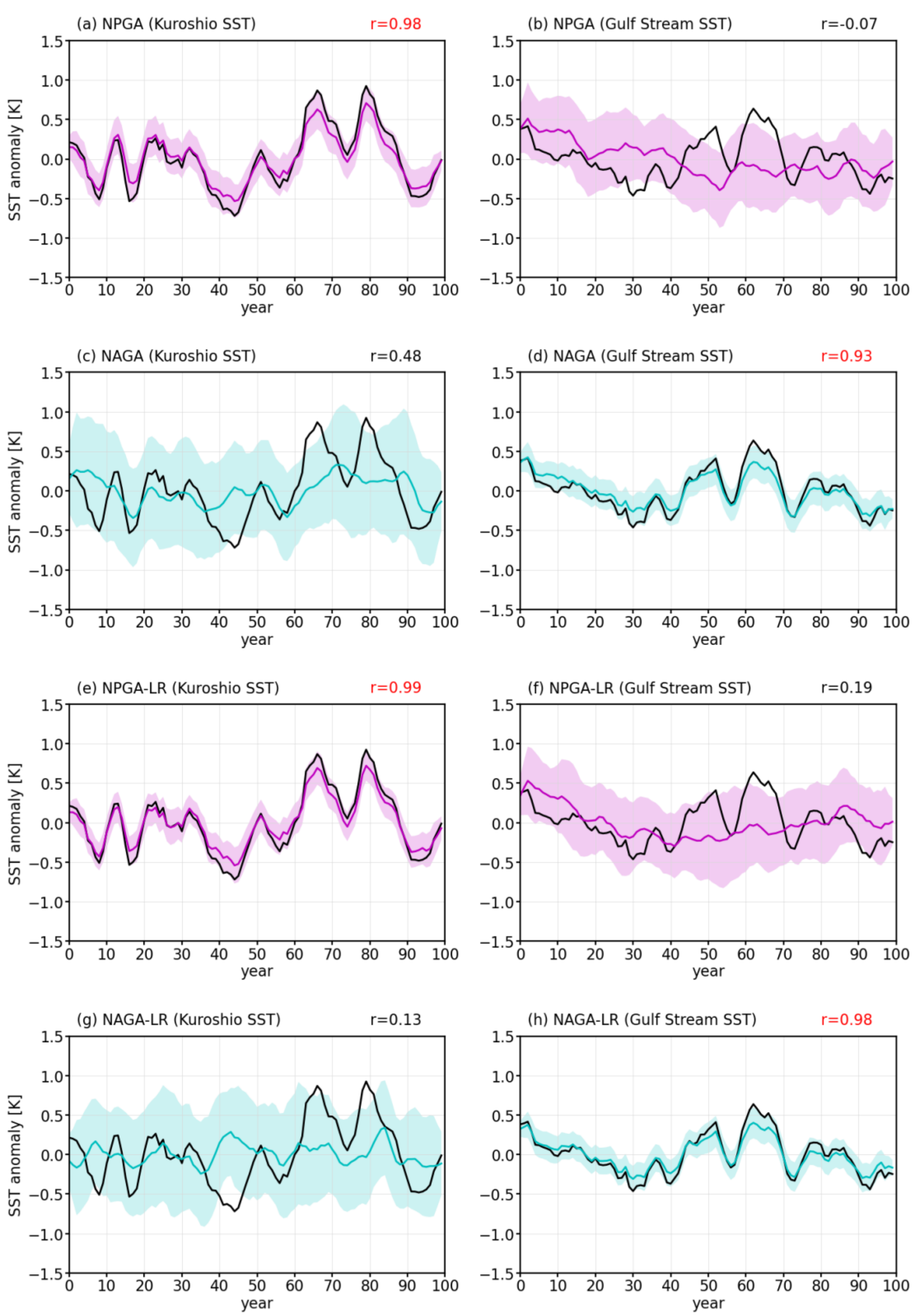

**Supplementary Figure 3. Simulated time series of the low-pass filtered Kuroshio and Gulf Stream SST anomalies.** As in Supplementary Fig. 2, but with five-year

running means.

**Supplementary Text 1**

**Statistical estimate of the variance fraction of Kuroshio SST**

To complement the dynamical experiments, we estimated the fraction of Kuroshio SST variance associated with Gulf Stream SST variability and tropical Pacific forcing using a simple linear regression model, following the statistical framework of Newman et al. (2016). In this framework, Kuroshio SST anomalies were represented as a function of its own persistence, Gulf Stream SST, and the Niño3.4 index;

$$Kuroshio(t) \ = \ a\, Kuroshio(t-1) \ + \ b\, Gulf\ Stream(t-4) \ + c\ Nino3.4(t) \ + \ d, \quad \text{(S1)}$$

where Kuroshio(t), Gulf Stream(t), and Nino3.4(t) denote the area-averaged annual Kuroshio SST, area-averaged annual Gulf Stream SST index, and DJF-mean Niño3.4 index, respectively. The Kuroshio(t-1) term represents first-order persistence (i.e., one-year before) in Kuroshio SST, whereas the Gulf Stream and Niño3.4 terms represent remote influences from the North Atlantic and tropical Pacific, respectively. We emphasize that the persistence term should not be interpreted as a direct estimate of local North Pacific forcing, because it may also include the memory of both local and remote processes.

The regression coefficients were estimated using annual-mean Kuroshio and Gulf Stream SST indices and the DJF-mean Niño3.4 index from the CTL simulation. Lead–lag correlation analyses were conducted to estimate the relevant lag time scales for the Gulf Stream SST and Niño3.4 indices (Supplementary Fig. 4a,b). Based on these analyses, we used Gulf Stream SST with a four-year lead relative to Kuroshio SST, (GS(t-4)), and Niño3.4 without any lag, (Nino3.4(t)). Because tropical Pacific indices at zero lag are commonly used in statistical models of Pacific decadal variability, such as Newman et al. (2016), we show the results based on the zero-lag Niño3.4 index in Supplementary Fig. 4. Note that we also tested Niño3.4 with a three-year lead relative to Kuroshio SST, but the results were qualitatively consistent.

To assess the relative importance of each index, we additionally constructed reduced regression models that include only the persistent term, Gulf Stream SST, or Niño3.4, respectively;

$$Kuroshio(t) \ = \ a\, Kuroshio(t-1) \ + \ d, \quad \text{(S2)}$$

$$Kuroshio(t) \ = \ b\, Gulf\ Stream(t-4) \ + \ d, \quad \text{(S3)}$$

$$Kuroshio(t) = c\,Nino3.4(t) + d. \text{(S4)}$$

The reconstruction based on the full statistical model using all predictors in Eq. (S1) captures Kuroshio SST variability reasonably well (Supplementary Fig. 4c). This result indicates that a large fraction of Kuroshio SST variability can be explained by persistence and further suggests that the remote influence from the Gulf Stream may make a non-negligible contribution.

We then calculated the explained variance for each regression model (Supplementary Fig. 4d). The full model explains 45% of Kuroshio SST variance, whereas the persistence-only model explains 41%. In the single-predictor models, Gulf Stream SST explains approximately 9% of Kuroshio SST variance, whereas Niño3.4 explains approximately 4%. Although these estimates should not be interpreted as an exact partitioning of local and remote contributions, they provide a statistical indication that Gulf Stream SST variability exerts a measurable remote influence on Kuroshio SST variability. The larger variance fraction associated with Gulf Stream SST than with Niño3.4 is also consistent with the relative contributions inferred from the NAGA and POGA experiments. These statistical estimates are also subject to uncertainty associated with the choice of lag used to represent remote forcings. Because remote influences can accumulate over timescales of up to about five years, this statistical model may underestimate their effects.

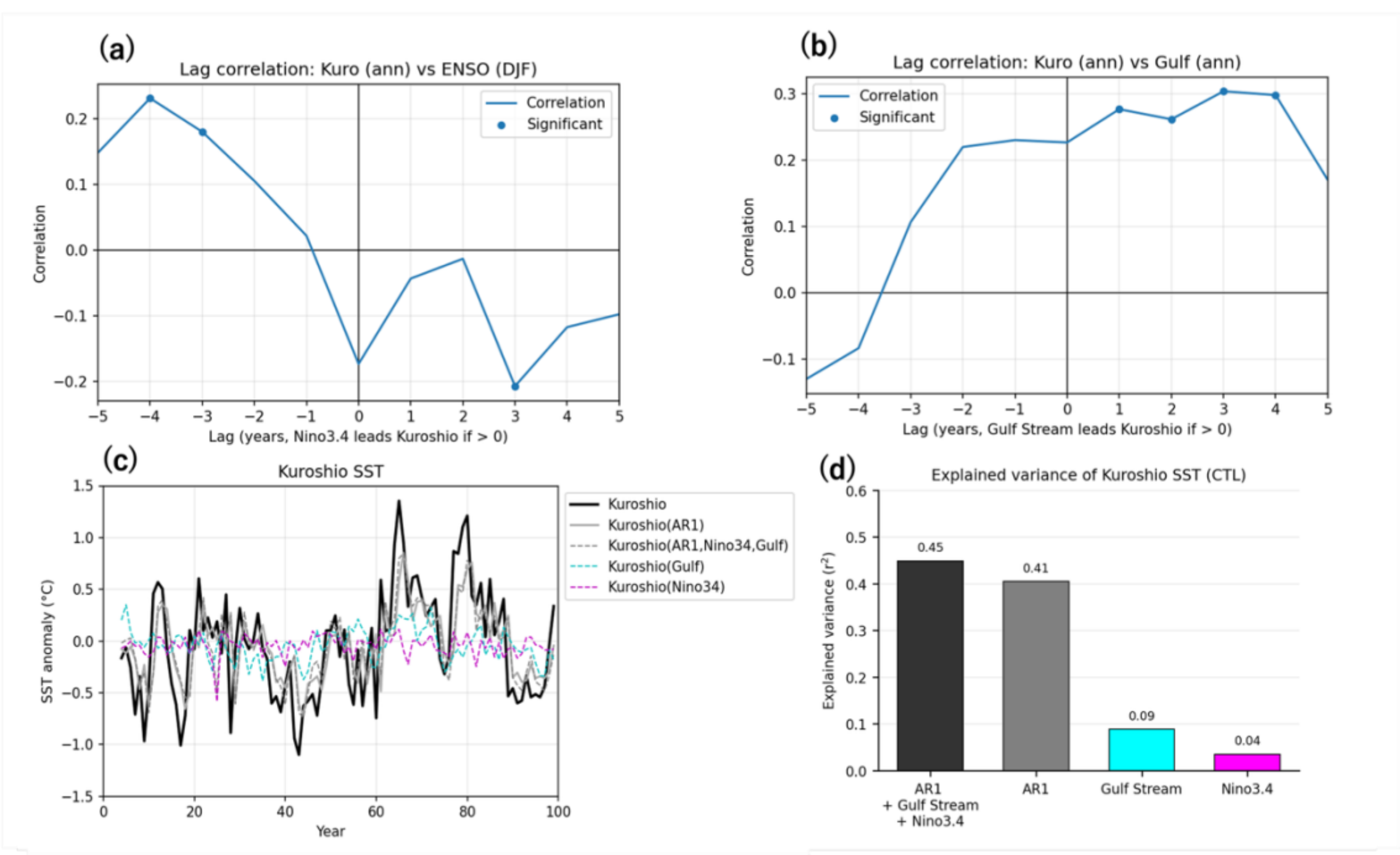

**Supplementary Figure 4. Lead–lag correlations, reconstructed Kuroshio SST time**

**series, and explained variance in CTL.** (a) Lead–lag correlation coefficients between annual-mean Kuroshio SST and the DJF-mean Niño3.4 index. The DJF mean is defined as the average of December of the preceding year and January–February of the current year. Positive lag indicates that Niño3.4 leads Kuroshio SST. Dots indicate statistically significant correlation coefficients. (b) As in (a), but for the correlation between annual-mean Kuroshio SST and annual-mean Gulf Stream SST. The correlations in (a) and (b) were calculated using time series that include linear trends. (c) Time series of Kuroshio SST in CTL (black), its reconstruction based on the full statistical model using all predictors (gray dashed; Eq. (S1)), and reconstructions based on single-predictor statistical models using persistence only (gray; Eq. (S2)), Gulf Stream SST only (cyan; Eq. (S3)), and Niño3.4 only (magenta; Eq. (S4)). (d) Explained variance of Kuroshio SST for each statistical model.

**Supplementary Text 2.**

**SVD and wavelet analysis**

A Singular Value Decomposition (SVD) analysis of ensemble-mean SST anomalies in the North Pacific (120°E-180°E, 25°N-55°N) and North Atlantic (30°W-90°W, 25°N-55°N) are conducted. The results show that NAGA captures in-phase co-variability of Kuroshio and Gulf Stream SSTs on decadal timescales (Supplementary Fig. 5). In CTL, the first SVD mode extracts coherent in-phase SST variations between the Kuroshio and Gulf Stream regions (r = 0.45), characterized by their meridional shifts. NAGA reproduces a similar in-phase co-variation as its first mode (r = 0.44), with prominent SST anomalies located north of the Kuroshio Extension and in the eastern part of the Gulf Stream Extension, although positive anomalies near the U.S. east coast are relatively weak. Conversely, both NPGA and NPGA-LR capture unclear SST patterns as their first SVD mode. While the first SVD mode in NAGA-LR also yields an in-phase co-variation in its leading SVD mode (r = 0.41), the spatial pattern is unclear in the Kuroshio region, underscoring the importance of eddy-permitting oceanic resolution in representing the influence of Gulf Stream variability on Kuroshio SSTs.

Wavelet analysis of SST anomalies provides additional support for the proposed causal relationship, demonstrating that constraining Gulf Stream SSTs reproduces low-frequency variations in Kuroshio SSTs, but not the reverse (Supplementary Fig. 6). In CTL, both the Kuroshio and Gulf Stream SSTs exhibit significant decadal-to-interdecadal variability. The interdecadal signature is reproduced in the ensemble-mean Kuroshio SST in NAGA but is absent from the Gulf Stream SST in NPGA (Supplementary Fig. 6c, d). Although NAGA-LR also shows significant interdecadal variability, this signal should be interpreted with caution. Given its inconsistency with the CTL time series (Fig. 1c; Extended Data Fig. 1) and the unclear spatial pattern in the SVD analysis, it is unlikely to represent a physically meaningful influence of Gulf Stream SST. Together, these results support a influence in which Gulf Stream SST variability modulates Kuroshio SST variability in NAGA. We therefore explore the physical mechanisms underlying the remote influence of the Gulf Stream on the Kuroshio, focusing primarily on the NAGA experiment.

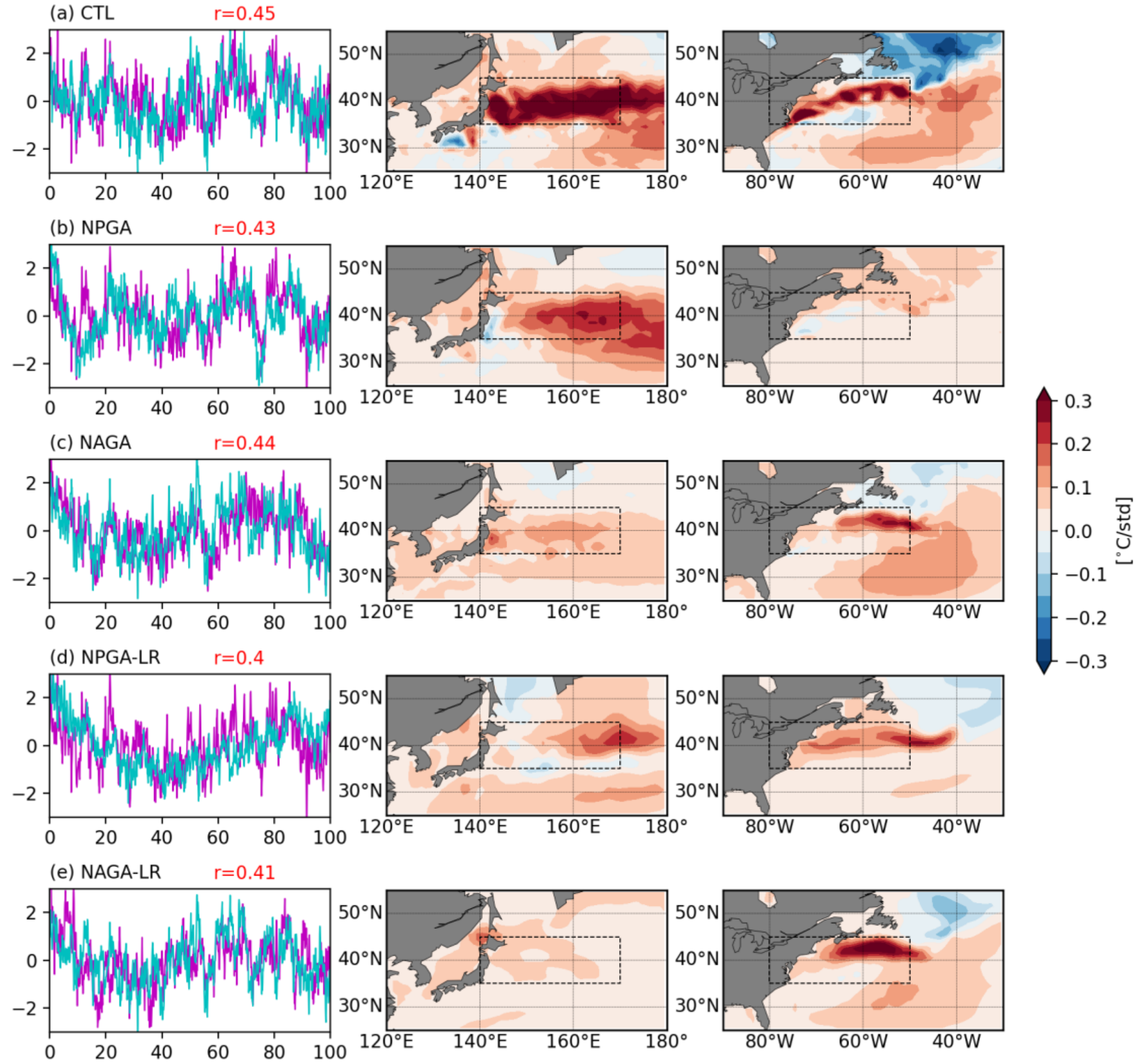

**Supplementary Figure 5. Dominant co-variability of Kuroshio and Gulf Stream SSTs.** (a) (Left) Monthly SST time series projected on the first mode of the SVD (SVD1) between SST anomalies in the North Pacific (25°N–55°N, 120°E–180°; magenta) and North Atlantic (25°N-55°N, 90°W-30°W; cyan) for CTL. The correlation coefficient between the two time series is shown; red indicates statistical significance at the 90% level. (Middle) SST anomaly pattern in the North Pacific regressed onto the normalized SVD1. (Right) SST anomaly pattern in the North Atlantic regressed onto the normalized SVD1. **(b–e)** As in **(a)**, but for **(b)** NPGA, **(c)** NAGA, **(d)** NPGA-LR, and **(e)** NAGA-LR.

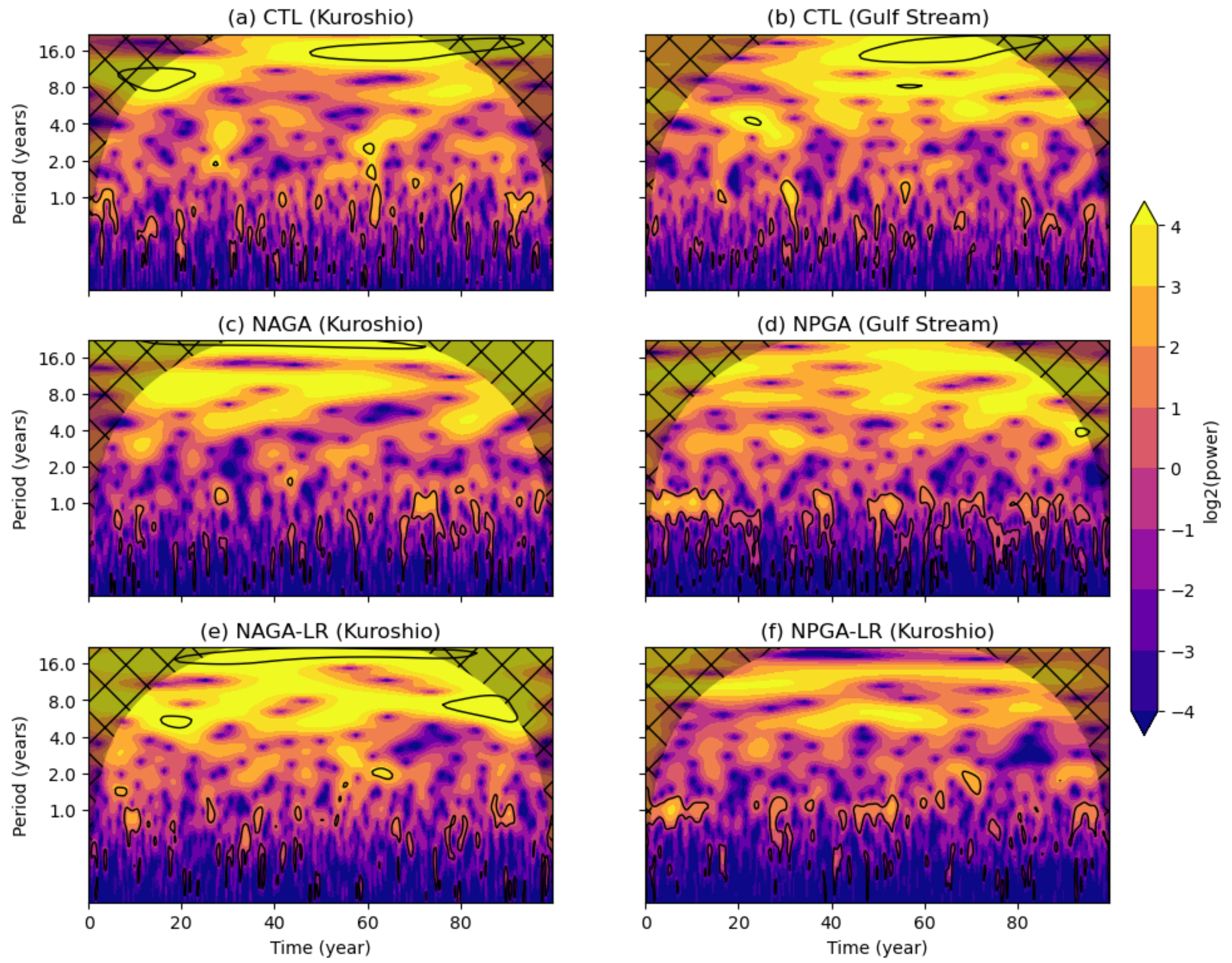

**Supplementary Figure 6. Normalized wavelet power spectra of SST anomalies in the Kuroshio and Gulf Stream regions.** (a, b) Normalized wavelet power spectra (shading) and 95% confidence level contour (black contour) for (a) Kuroshio and (b) Gulf Stream SST anomalies in CTL. The vertical axis represents the period in years (logarithmic scale), while the horizontal axis indicates the simulation year. Hatching indicates the cone of influence. **(c, d)** As in **(a, b)**, but for the ensemble means of **(c)** Kuroshio SST anomalies in NAGA and **(d)** Gulf Stream SST anomalies in NPGA. **(e, f)** As in **(c, d)**, but for **(e)** Kuroshio SST anomalies in NAGA-LR and **(f)** Gulf Stream SST anomalies in NPGA-LR.

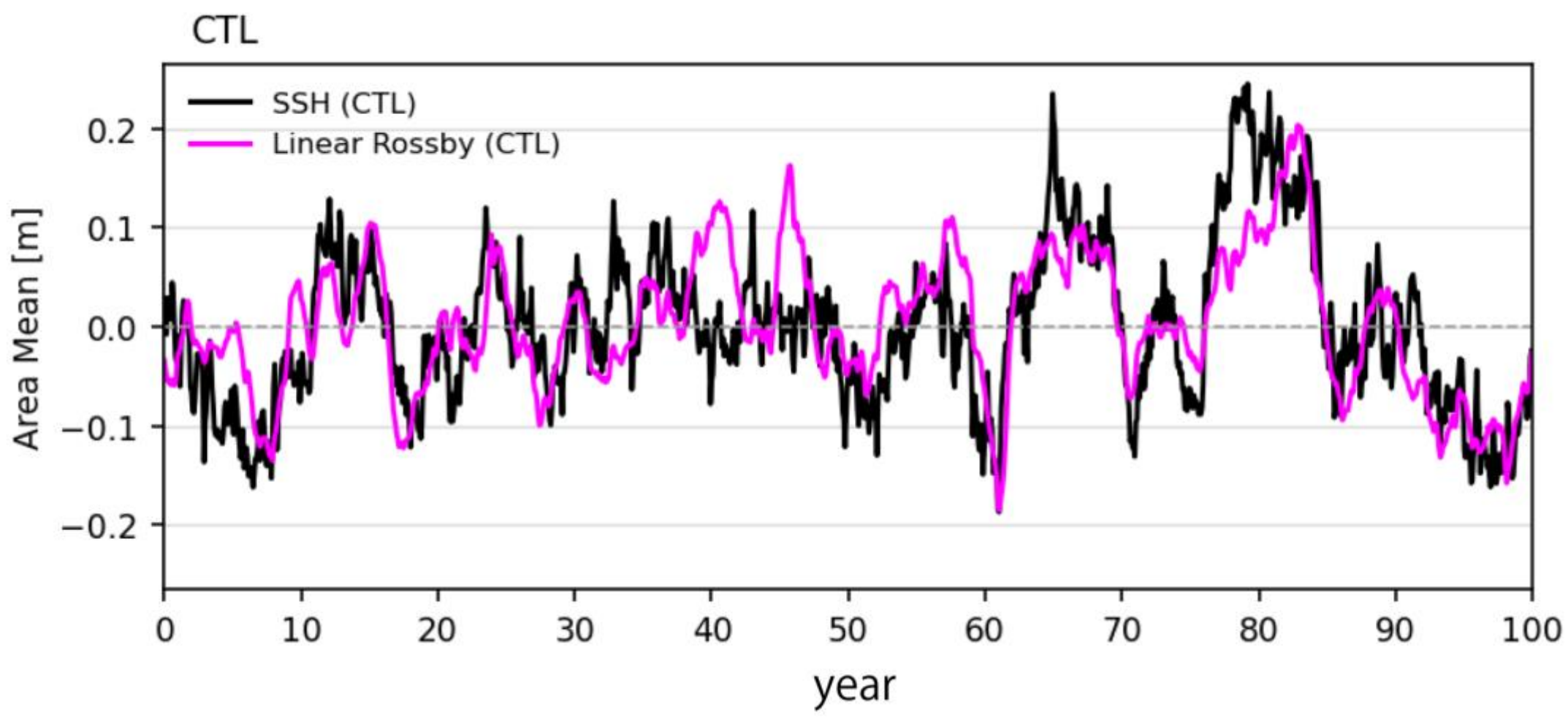

**Supplementary Figure 7. Comparison of SSH anomalies and the reconstruction by linear Rossby wave model in the Kuroshio Extension region.** Monthly SSH anomalies [m] for CTL (black) and linear Rossby wave model (pink) averaged over the Kuroshio Extension (140°E–170°E, 31°N–36°N).

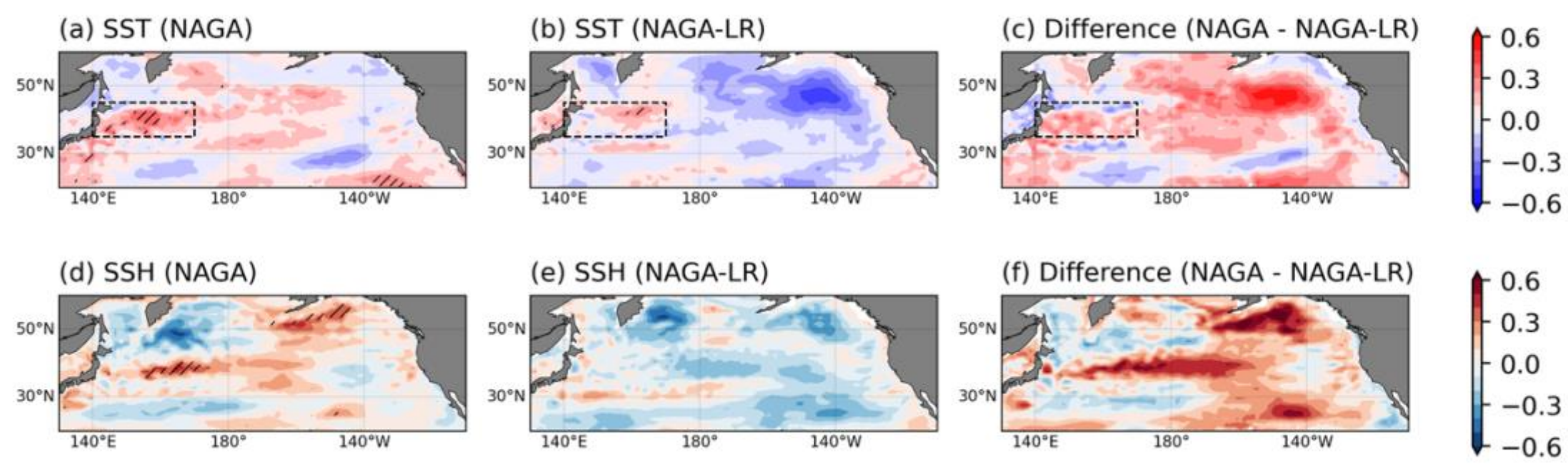

**Supplementary Figure 8. Reproduction of SST and SSH anomalies in NAGA and NAGA-LR. (a)** Pointwise correlation coefficients of annual SST anomalies between CTL and the ensemble mean of NAGA. Hatching indicates statistically significant correlation at the 90% confidence level. Black dashed box indicates the Kuroshio region as defined in this study (140°E–170°E, 35°N–45°N). **(b)** As in **(a)**, but for NAGA-LR. **(c)** Difference in correlation coefficients between NAGA and NAGA-LR. **(d–f)** As in **(a–c)**, but for SSH anomalies.

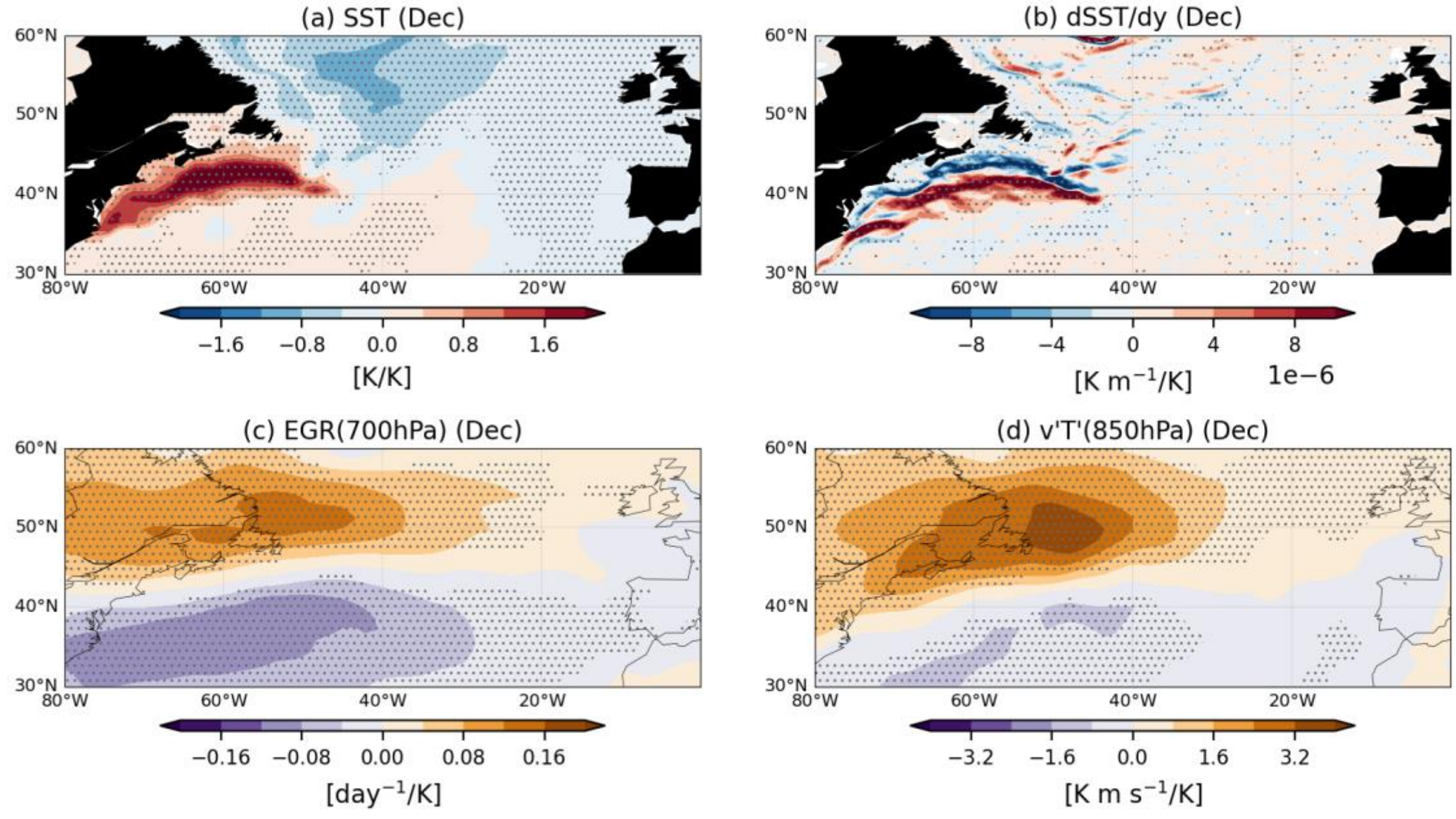

**Supplementary Figure 9. Response of surface and tropospheric North Atlantic circulation anomalies associated with Gulf Stream SST anomalies.** Regression coefficients for boreal early winter (December) SST, meridional SST gradient (dSST/dy), Eady growth rate (EGR) (Methods), storm-track activity at 850hPa for NAGA ensembles. All anomalies regressed onto the DJF mean SST anomalies in the Gulf Stream region. Stippling indicates regression coefficients statistically significant at the 90% confidence level.

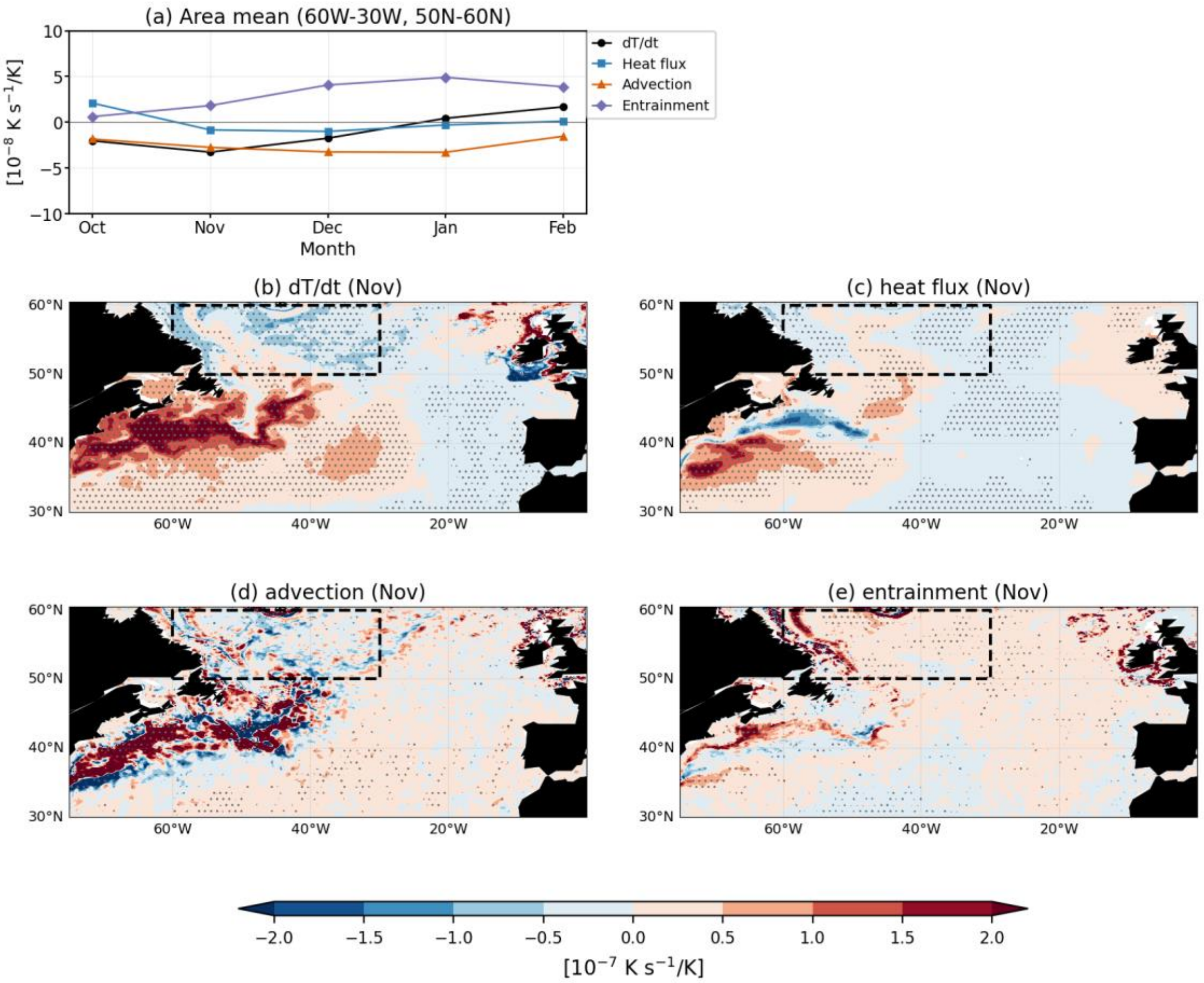

**Supplementary Figure 10. Simulated surface-ocean responses of the North Atlantic subpolar gyre associated with Gulf Stream SST warming.** (a) Temporal evolution of the monthly mixed-layer temperature tendency, surface heat flux contribution, and oceanic contribution, including horizontal advection and entrainment, to the mixed-layer temperature tendency within the subpolar gyre. All variables are area-averaged lag-regression coefficients onto DJF-mean Gulf Stream SST in NAGA, and thus have units of [$10^{-8}$ K $s^{-1}$ $K^{-1}$]. (b–e) Spatial patterns of each term regressed onto Gulf Stream SST anomalies in November when the negative mixed-layer temperature tendency peaks. (b) Mixed-layer temperature tendency, (c) surface heat flux, (d) horizontal advection, and (e) entrainment. Units are [$10^{-7}$ K $s^{-1}$ $K^{-1}$]. Hatching indicates statistically significant regression coefficients. The area average in (a) is taken over the black dashed boxes (60°W–30°W, 50°N–60°N).

**Supplementary Text 3.**

**Tropical-damping experiments**

To further assess the robustness of the Gulf Stream SST influence via mid-latitude atmospheric pathways, we conducted additional sensitivity experiments in which tropical SST anomalies were restored to climatological values, thereby supressing the impact of tropical SST anomalies associated with the Gulf Stream SST (Supplementary Table 2). Three experiments were performed by restoring SST to the CTL climatology over (1) the northern tropical Atlantic (CTL-NoATL), (2) the tropical Indian Ocean and Pacific Ocean (CTL-NoINPAC), and (3) the entire tropical Ocean (CTL-NoTRO) (Supplementary Fig. 11). All simulations were conducted as one-member preindustrial control-type simulations without constraining SST anomalies in the Gulf Stream region. All tropical SST-damping experiments are conducted at 0.25° oceanic horizontal resolution (i.e., MIROC6subhires) for 100 years. Because the results are qualitatively similar across all experiments, only the CTL-NoTRO results are shown hereafter.

All tropical-damping experiments preserve the pronounced co-variability between Kuroshio and Gulf Stream SST anomalies (Supplementary Fig.12) and reproduce the weakening of the Aleutian Low in association with Gulf Stream SST anomalies (Supplementary Fig. 13). Although these experiments exhibit a reduction in the amplitude and spatial coherence of the storm-track response, the persistence of the Aleutian Low weakening confirms that primary driver of the response in CTL and NAGA is extratropical SST forcing, tropical SST anomalies playing a secondary role.

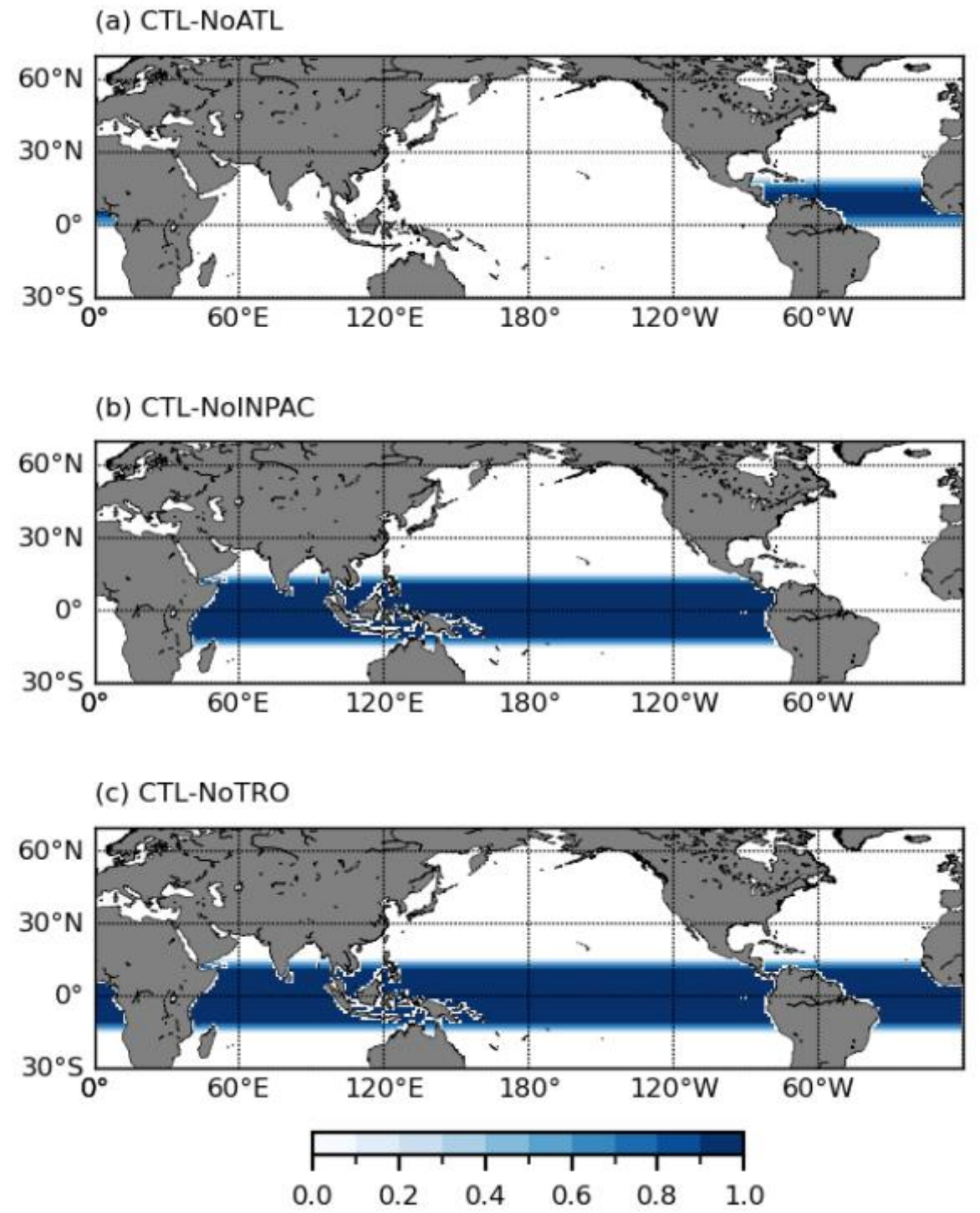

**Supplementary Figure 11. Nudging regions for tropical SST-damping experiments. (a)** As in Supplementary Fig. 1, but showing the nudging heat flux mask for CTL-NoATL. Colours indicates regions where SST is restored to the CTL climatology. As in **(a)**, but for **(b)** CTL-NoINPAC, and **(c)** CTL-NoTRO.

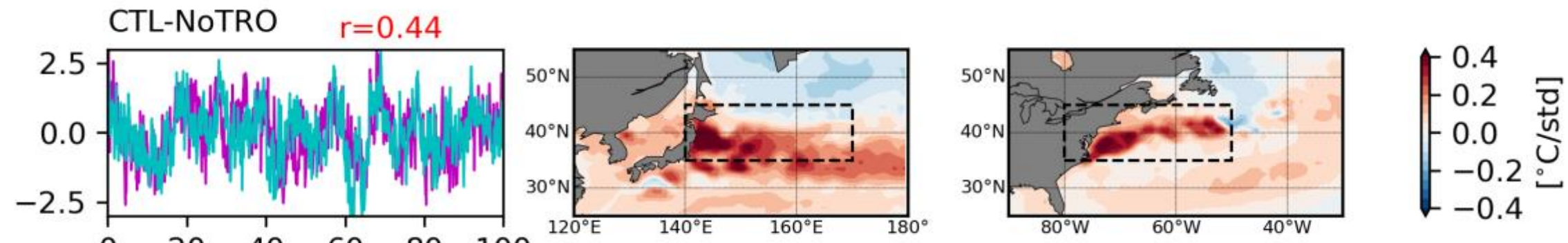

**Supplementary Figure 12. SVD analysis for tropical SST-damping experiments.** As in Supplementary Figure 5, but applied to the tropical SST-damping experiments (CTL-NoTRO).

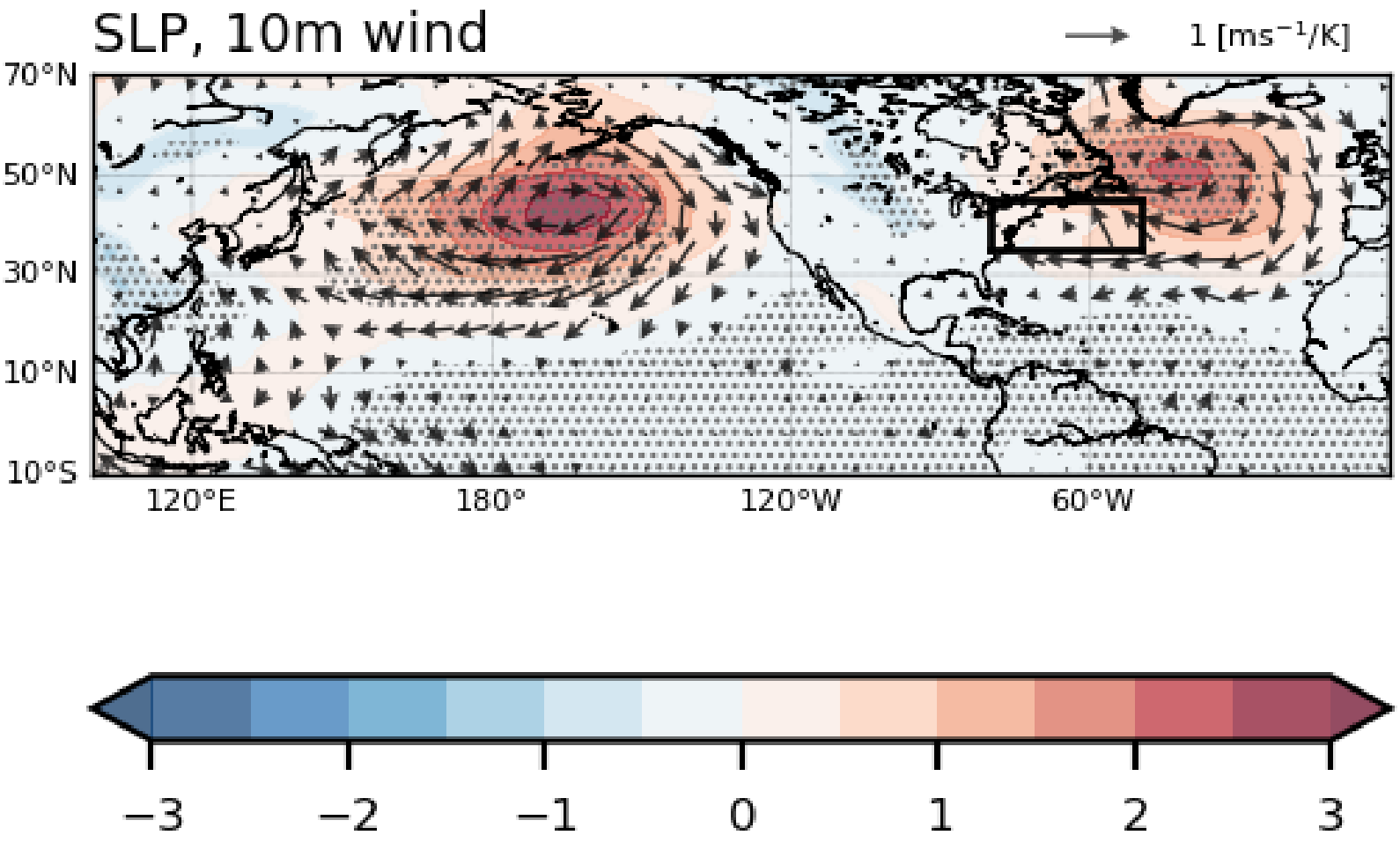

**Supplementary Figure 13. Response of surface and tropospheric circulation anomalies associated with Gulf Stream SST anomalies.** Regression coefficients for boreal winter (DJF: December–January–February) mean sea level pressure (SLP; shading) [hPa/K] and 10 m wind (vectors) [$ms^{-1}$/K] regressed onto the DJF mean SST anomalies in the Gulf Stream region for CTL-NoTro. Stippling indicates regression coefficients statistically significant at the 90% confidence level.

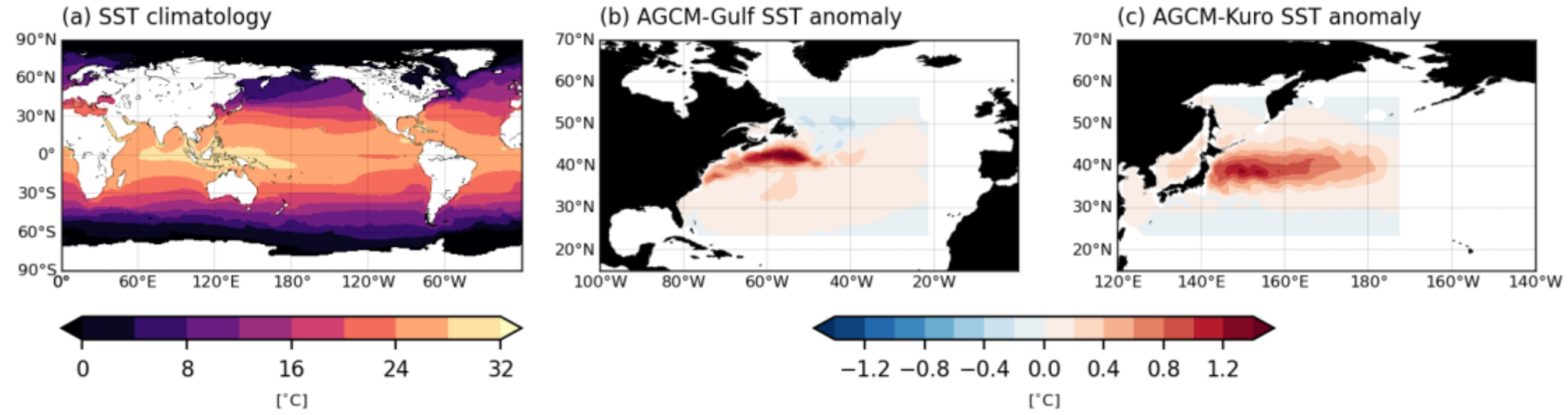

**Supplementary Figure 14. SST anomalies prescribed in the AGCM experiments. (a)** Annual mean climatological SST prescribed in AGCM-Clim, calculated from CTL. (**b, c)** SST anomalies prescribed in **(b)** AGCM_Gulf and **(c)** AGCM_Kuro. Anomalies are defined as regression coefficients of SST anomalies onto Gulf Stream SST and Kuroshio SST anomalies in CTL, respectively. The unit is [°C], representing SST anomalies comparable to those when the Gulf Stream and Kuroshio SST are +1°C, respectively.

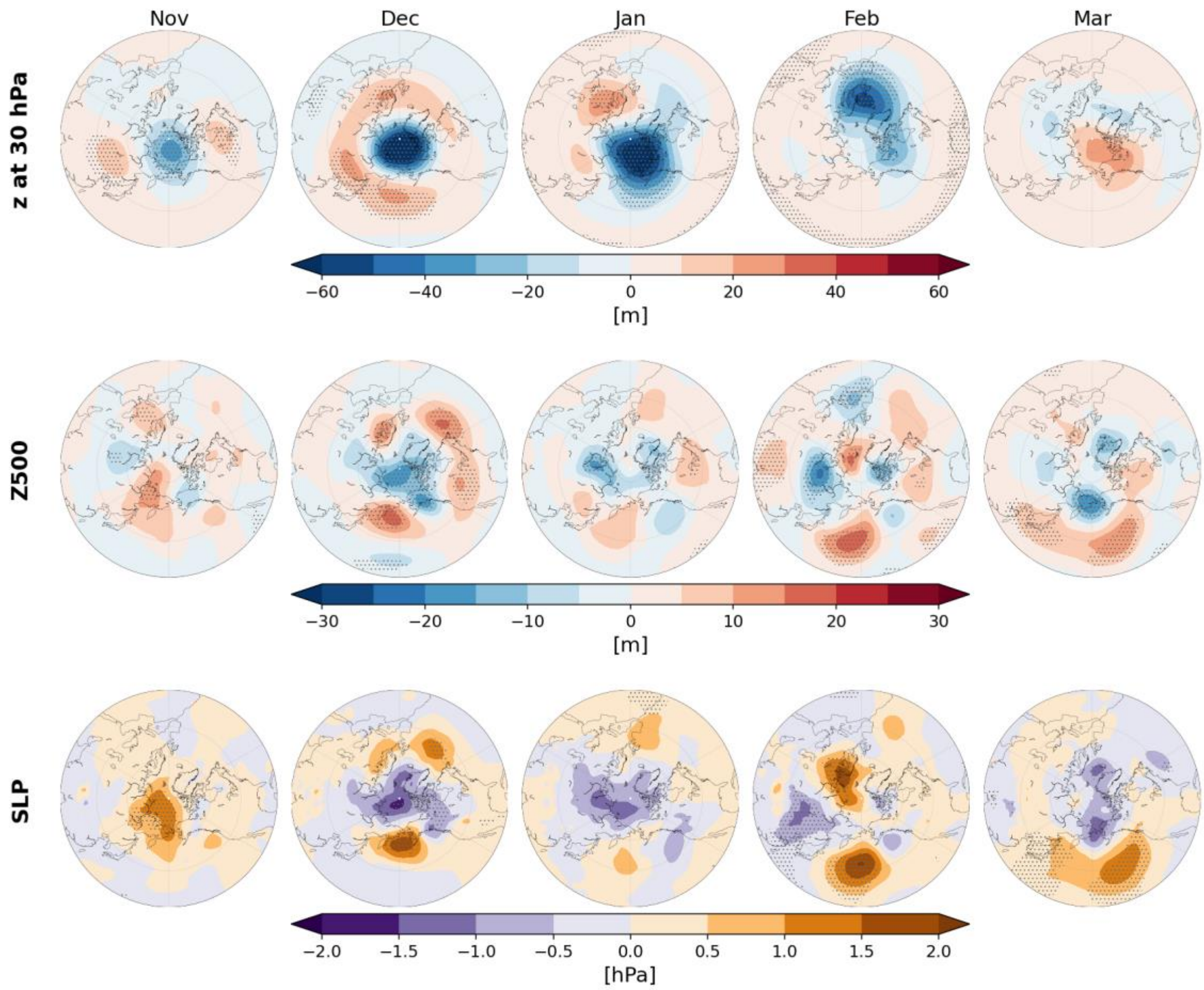

**Supplementary Figure 15. Response of atmospheric circulation anomalies associated with the Gulf Stream SST anomalies in AGCM experiments.** As in Extended Data 5, but showing differences between AGCM-Gulf_SST and AGCM-Clim.

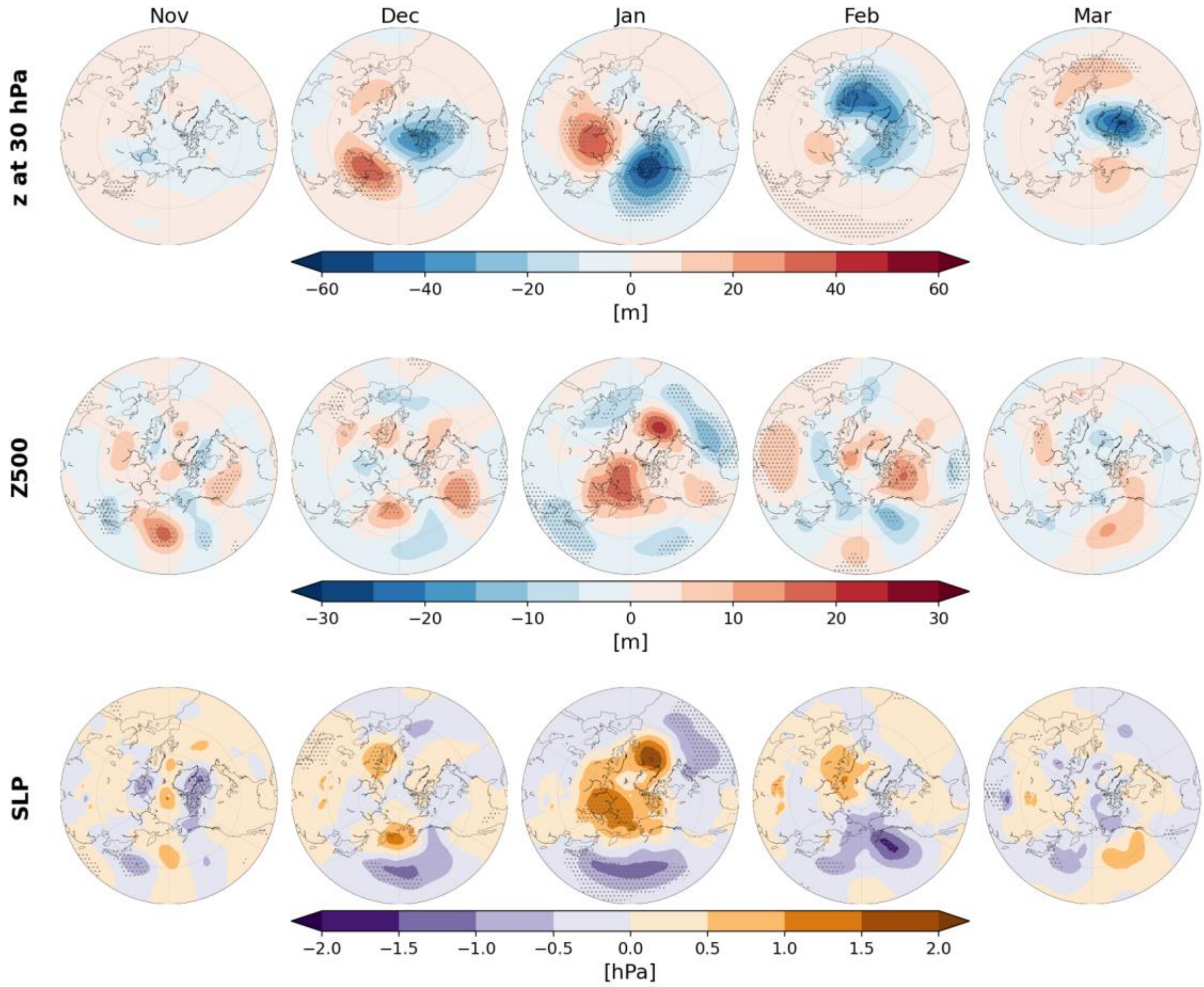

**Supplementary Figure 16. Response of atmospheric circulation anomalies associated with the Kuroshio SST anomalies in AGCM experiments.** As in Supplementary Figure 15, but for differences between AGCM_Kuro and AGCM_Clim.

**Supplementary Text 4**

**Diagnose of Eliassen-Palm flux for AGCM experiments**

To examine the influence of Gulf Stream SST anomalies on the stratosphere, we calculated differences in Eliassen–Palm (E–P) flux between AGCM_Gulf and AGCM_Clim. The E-P flux was formulated in pressure coordinate on the sphere following the previous study (Andrew et al. 1983; see Methods). This comparison allows us to diagnose planetary-wave anomalies and the associated stratospheric response attributable to Gulf Stream SST anomalies. Because AGCM_Gulf exhibits a strengthening of the stratospheric polar vortex in early winter, consistent with NAGA (Supplementary Fig. 15), we focused on anomalies during November–December.

The E–P flux forcing indicates stratospheric westerly acceleration at 50°N–80°N and westerly deceleration at 20°N–50°N (Supplementary Fig. 17). This pattern appears to arise from an equatorward distortion of E–P flux propagation, particularly in the stratosphere. Although a more detailed investigation of the underlying mechanism is beyond the scope of this study, these results support the possibility that Gulf Stream SST anomalies induce an early-winter stratospheric response through anomalies in planetary-wave propagation. This interpretation is consistent with a previous study suggesting that meridional shifts of the North Atlantic storm track can distort planetary-wave propagation (Limpasuvan and Hartmann 2000).

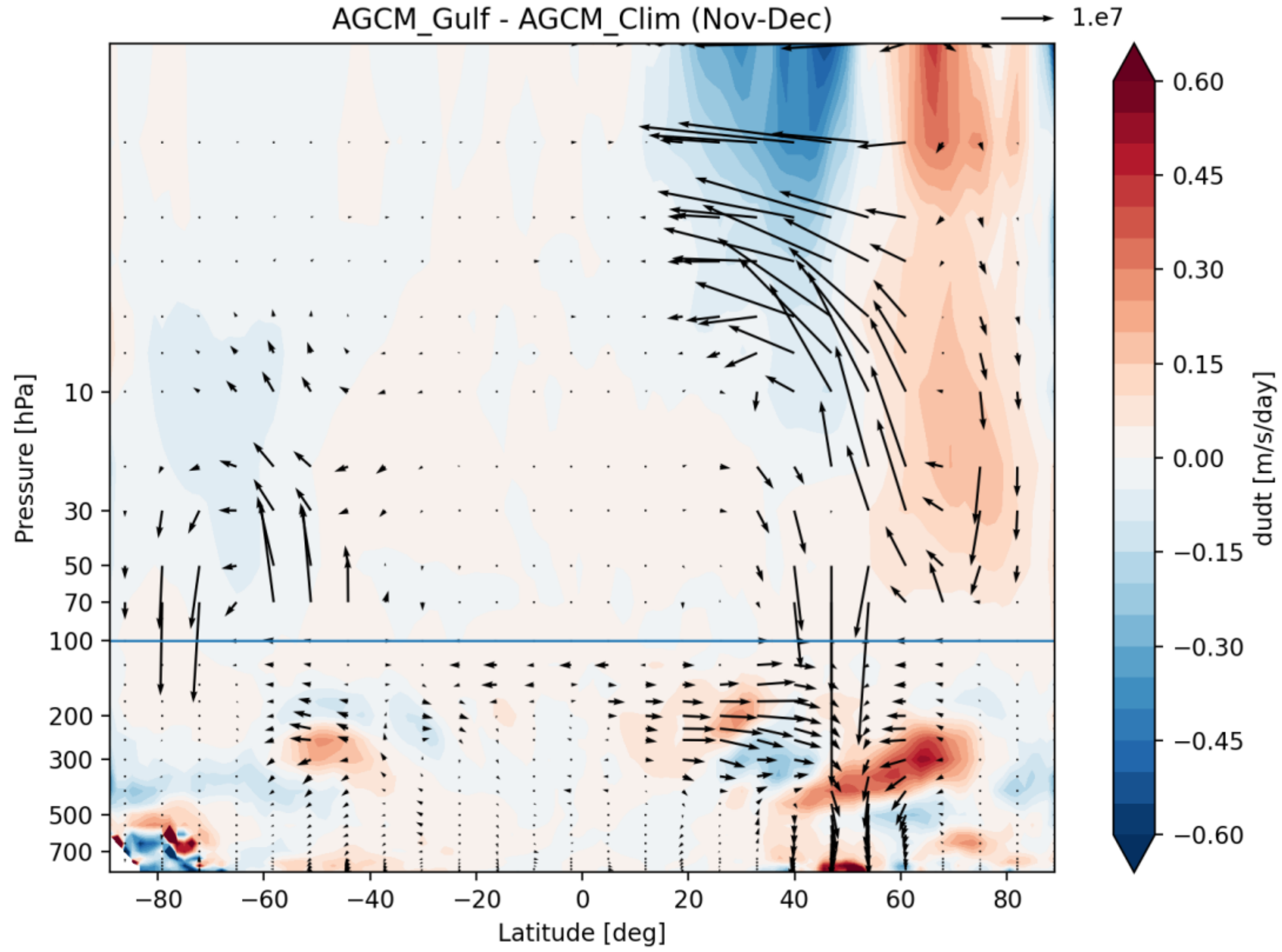

**Supplementary Figure 17 Differences in E–P flux and its westerly-wind acceleration associated with Gulf Stream SST anomalies in AGCM_Gulf during early winter.** Differences in E–P flux (vectors) and zonal-mean zonal wind anomalies induced by E–P flux divergence [m $s^{-1}$ $day^{-1}$] (shading) are shown for the November–December mean. Both fields are shown as differences between AGCM_Gulf and AGCM_Clim. For visualization, only the vertical component of the E–P flux vectors above 100 hPa is multiplied by 100.

**Supplementary Text 5**

**Atmospheric responses in NAGA-LR**

In NAGA-LR, the weakening of the Aleutian Low is less distinct (Supplementary Fig. 18). The meridional dipole structure over the North Atlantic is weaker than that in CTL, and, in contrast to NAGA, the Aleutian Low strengthens rather than weakens. This resolution-dependent contrast suggests that, in NAGA-LR, the associated wind stress anomalies are too weak to effectively influence the Kuroshio Extension. Furthermore, SST anomalies in the North Atlantic subpolar gyre are unclear, particularly over the Labrador Sea. This implies that the coarse ocean resolution of NAGA-LR prevents it from representing the meridional shift of the Gulf Stream SST front, as well as the associated atmospheric response and subpolar SST variability.

The low-resolution control simulation, MIROC6 (CTL-LR), represents the meridional SST dipole and the NAO-like atmospheric circulation pattern over the North Atlantic somewhat better than HighResMIP models with comparable non-eddying ocean resolutions (Extended Data Fig. 8). Nevertheless, these structures remain weaker than those simulated by models with eddy-permitting ocean resolutions (Extended Data Fig. 8c), suggesting that the representation of atmospheric responses to Gulf Stream SST variability in CTL-LR is limited by the coarse ocean resolution, as represented in NAGA-LR.

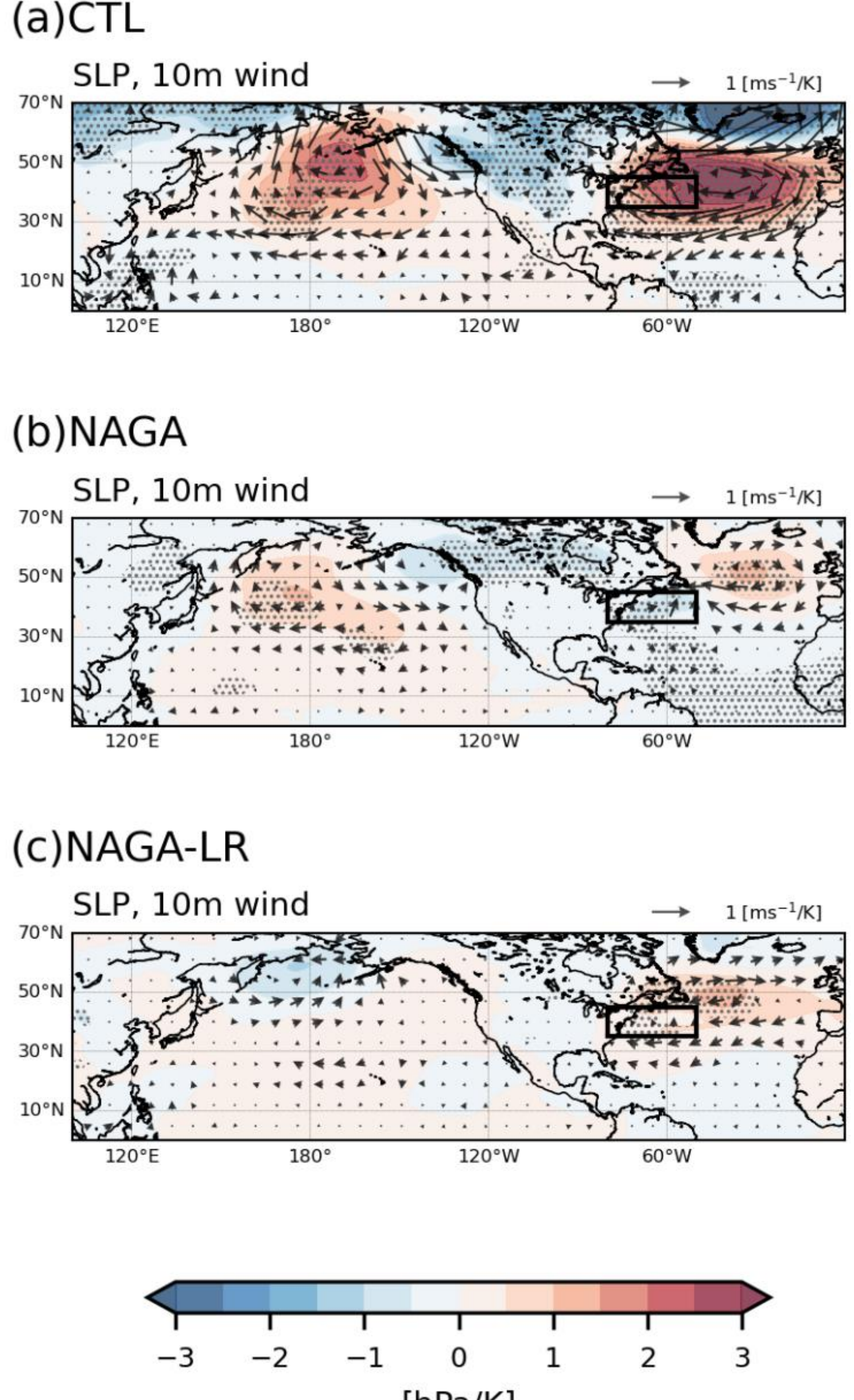

**Supplementary Figure 18 Response of surface and tropospheric circulation anomalies associated with Gulf Stream SST anomalies. (a)** Regression coefficients for boreal winter (DJF: December–January–February) mean sea level pressure (SLP; shading) [hPa/K] and 10 m wind (vectors) [$ms^{-1}$/K] regressed onto the DJF mean SST anomalies in the Gulf Stream region for CTL. Stippling indicates regression coefficients statistically significant at the 90% confidence level. (**b–c)** As in **(a)**, but for **(b)** NAGA and **(c)** NAGA-LR.

**Supplementary Text 6**

**Observational evidence of atmospheric responses to Gulf Stream and Kuroshio SST anomalies.**

We further investigated the relationships between Gulf Stream and Kuroshio sea surface temperature (SST) variability and atmospheric circulation anomalies using ERA5 reanalysis and observational SST dataset (COBE-SST2). The findings provide observational evidence that is broadly consistent with the atmospheric responses to Gulf Stream SST anomalies that were identified in the pacemaker and atmospheric general circulation model (AGCM) experiments.

Regression analyses based on Gulf Stream SST indicate that positive Gulf Stream SST anomalies are associated with North Atlantic Oscillation (NAO)-like circulation anomalies throughout the North Atlantic troposphere and at the surface during early winter, accompanied by a strengthening of the stratospheric polar vortex. This response is subsequently followed by a weakening of the Aleutian Low in late winter (Supplementary Fig. 19).

We further examined the stratosphere–troposphere coupling associated with the Northern Annular Mode (NAM)-like response. As seen in the NAGA and AGCM_Gulf experiments, Gulf Stream SST anomalies are associated with a strengthening of the polar vortex in early winter and subsequent downward-propagating anomalies, although the signal is less robust, particularly with respect to the downward influence on the troposphere (Supplementary Fig. 20). This uncertainty is likely attributable to the limited number of events identified in the composite analysis, with only nine cases available, thereby restricting the extent to which internal variability can be effectively isolated.

The atmospheric response to Kuroshio SST variability is characterized by a weakening of the Aleutian Low in early winter and a strengthened stratospheric polar vortex in late winter (Supplementary Fig. 21). However, as seen in NPGA, NAO-like anomalies are not formed over the North Atlantic, indicating that the North Atlantic atmospheric response to Kuroshio SST variability is unclear. At the same time, these observational results should be interpreted with caution, given the difficulty of distinguishing forced responses from internal atmospheric variability over the relatively short observational record and of establishing causality in air–sea interactions. Overall, these findings suggest that the atmospheric response to Kuroshio Extension SST variability is less robustly identified than the response to Gulf Stream SST variability, highlighting the need for additional observational constraints and continued model development.

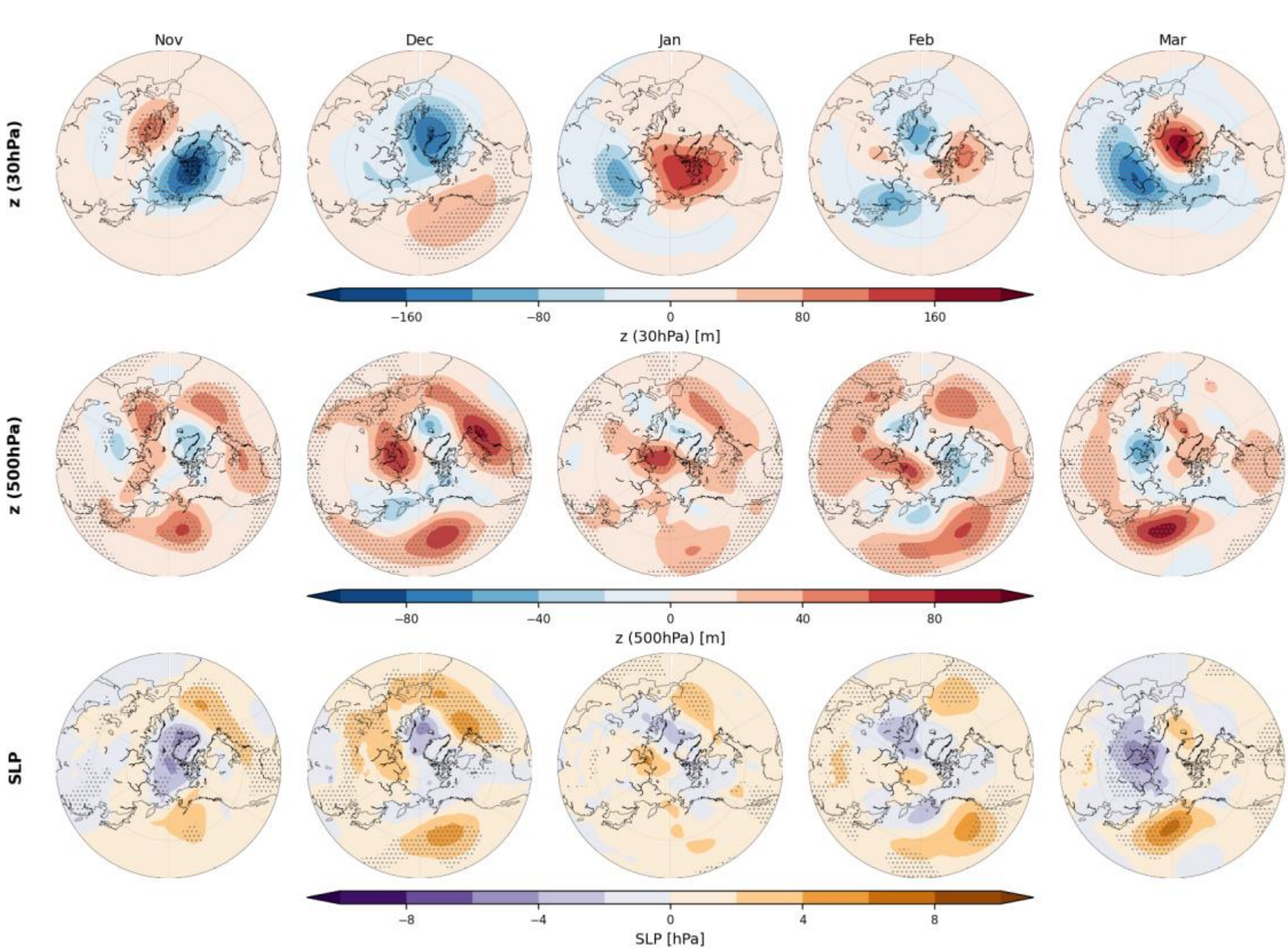

**Supplementary Figure 19 Response of tropospheric and stratospheric circulation anomalies associated with Gulf Stream SST anomalies.** Regression coefficients of monthly mean geopotential height at 30 hPa (top), at 500hPa (middle), and sea level pressure (SLP; bottom) during boreal winter (from November to March) for ERA5 during 1980-2024. All variables are regressed to the DJF-mean SST (COBE-SST2) anomalies in the Gulf Stream region. Stippling indicates regression coefficients statistically significant at the 90% confidence level.

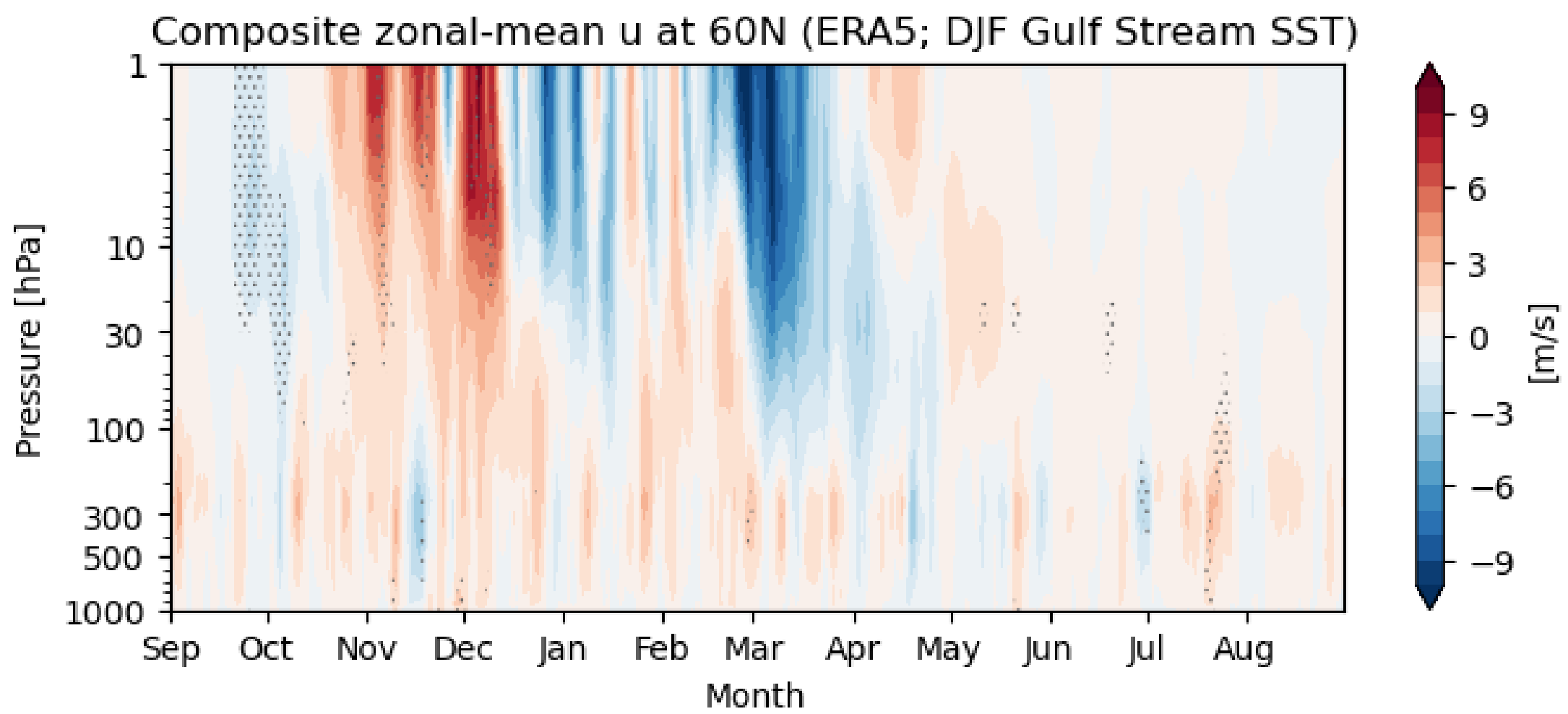

**Supplementary Figure 20. Response of tropospheric and stratospheric circulation anomalies associated with Gulf Stream SST anomalies.** Composites of zonal-mean zonal wind anomalies at 60°N when DJF Gulf Stream SST exceeds +1 standard deviation for ERA5 and COBE-SST2. Stippling indicates regression coefficients statistically significant at the 90% confidence level.

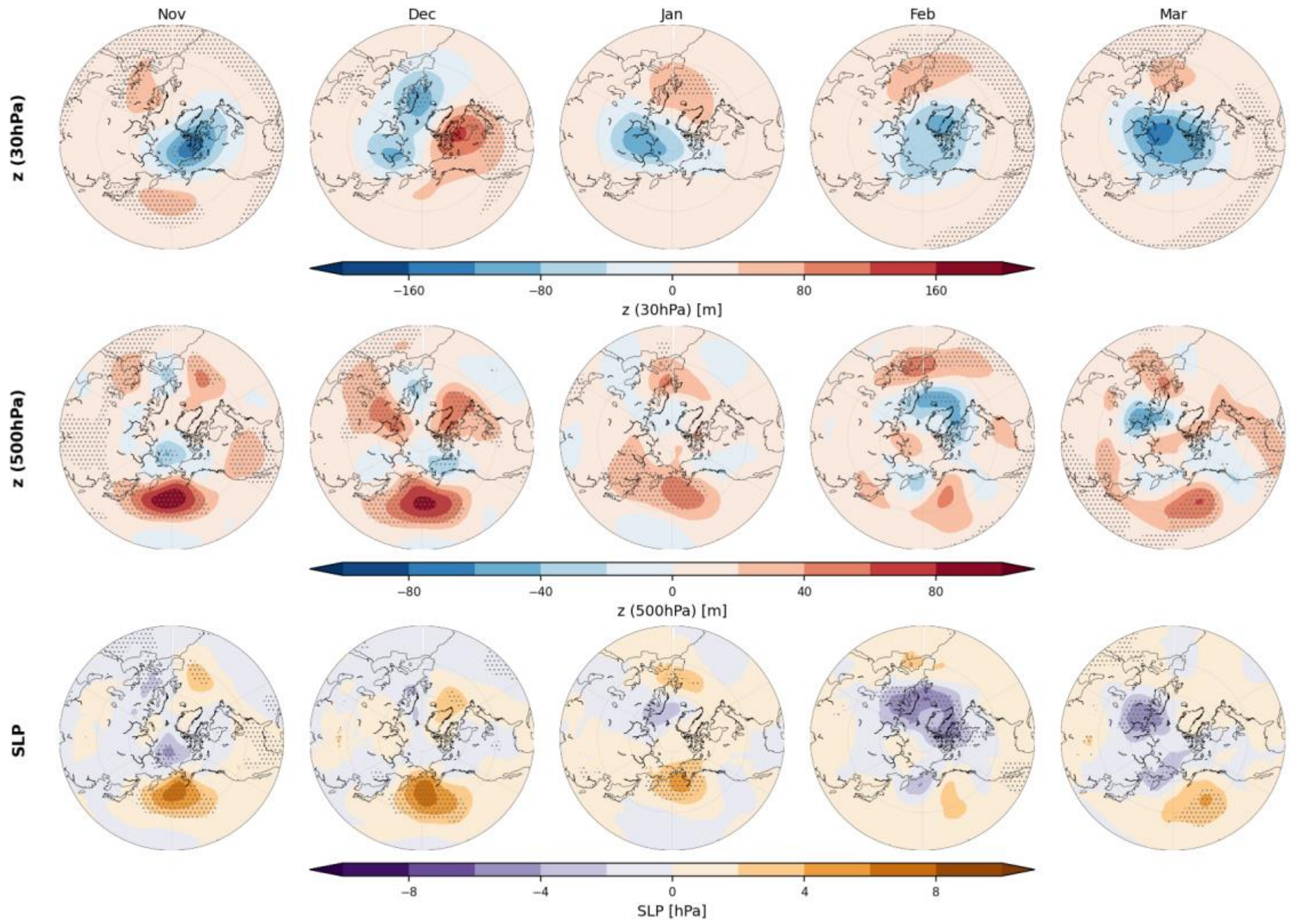

**Supplementary Figure 21 Response of tropospheric and stratospheric circulation anomalies associated with Kuroshio SST anomalies.** As in Supplementary Figure 19, but for regressions onto Kuroshio SST anomalies.

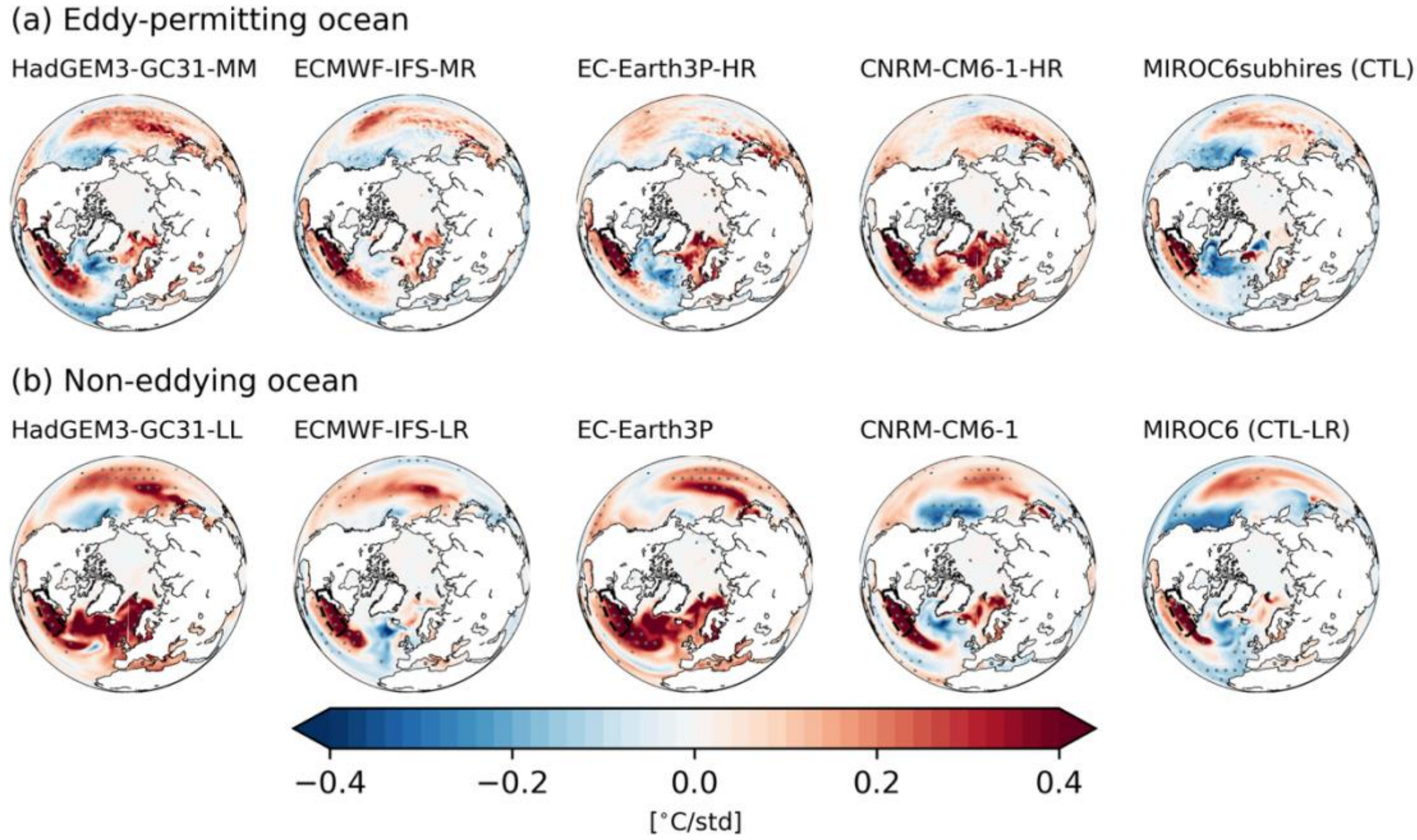

**Supplementary Figure 22. Resolution dependence of SST response to Gulf Stream SST anomalies in CMIP6 models.** DJF mean SST anomalies [°C/std] regressed onto normalized DJF mean SST anomalies in the Gulf Stream region (black dashed box) for CMIP6 models with **(a)** eddy-permitting and **(b)** non-eddying ocean resolution. Dots indicate statistically significant regression coefficients at the 90% confidence level.

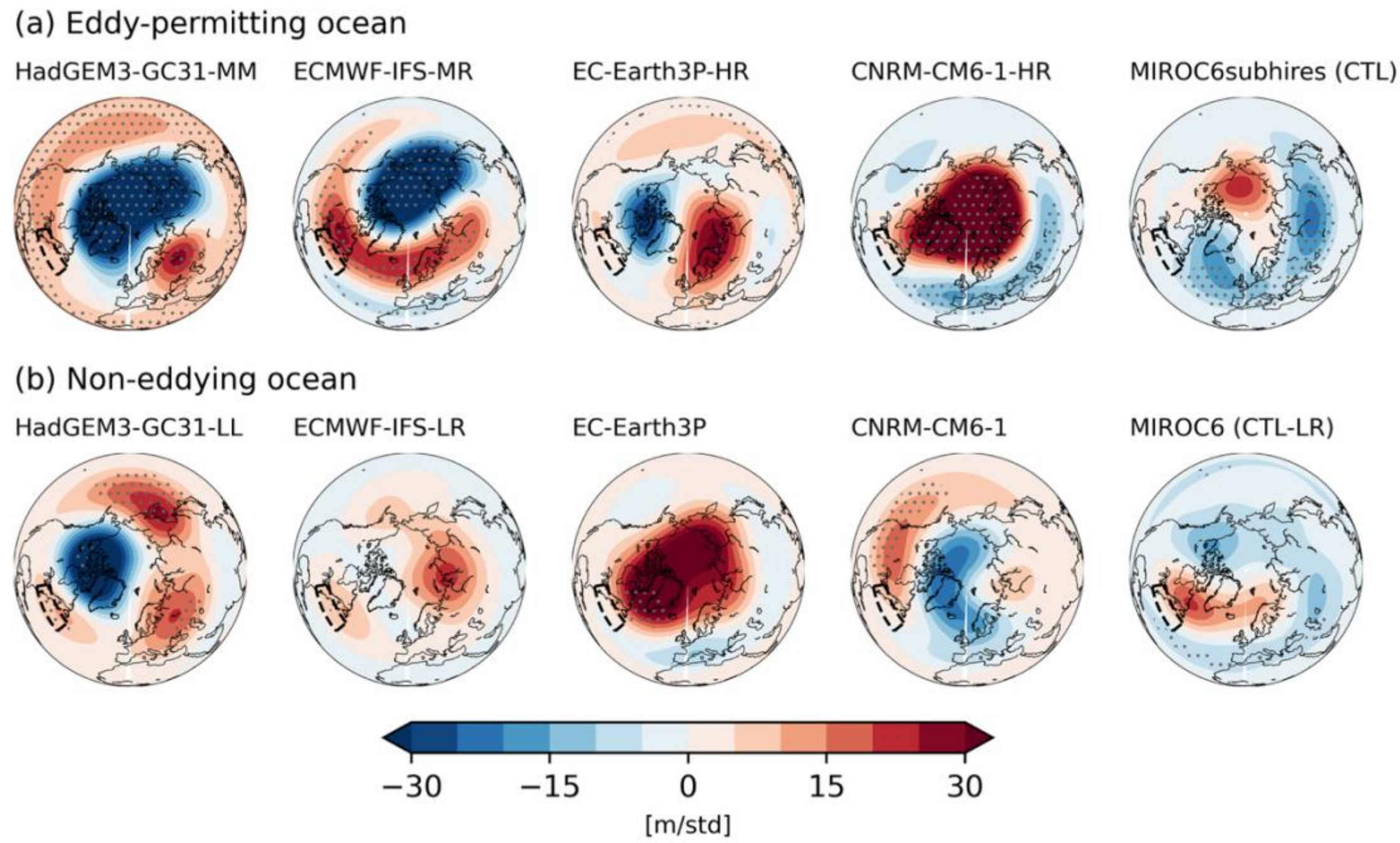

**Supplementary Figure 23. Resolution dependence of the stratospheric circulation response to the Gulf Stream SST anomalies in CMIP6 models.** DJF mean geopotential height at 30 hPa [m/std] regressed onto normalized DJF mean SST anomalies in the Gulf Stream region (black dashed box) for CMIP6 models with **(a)** eddy-permitting and **(b)** non-eddying ocean resolution. Dots indicate statistically significant regression coefficients at the 90% confidence level.

**References in Supplementary Data**